%% file: main.tex
\documentclass[a4paper,fleqn]{cas-sc}

\usepackage[numbers]{natbib}
\usepackage{siunitx} 
\usepackage{xtab} 
\usepackage{parnotes} 
\usepackage{caption} 
\usepackage{subcaption} 
\usepackage{breakcites}  
\usepackage{todonotes} 
\usepackage{longtable}  
\usepackage{enumitem}

\usepackage{graphicx} 
\graphicspath{{figures/}} 

\PassOptionsToPackage{hyphens}{url}
\usepackage{hyperref}
\usepackage[noabbrev,capitalize,nameinlink]{cleveref}

\DeclareSIUnit{\EUR}{\text{€}}
\DeclareSIUnit{\perkWh}{\text{\ensuremath{\text{€/kWh}}}}

\makeatletter
\g@addto@macro{\UrlBreaks}{\UrlOrds}
\g@addto@macro{\UrlNoBreaks}{\do\.} 
\makeatother

\usepackage[%
  newcommands     
 ,newparameters   
]{ragged2e}

\def\tsc#1{\csdef{#1}{\textsc{\lowercase{#1}}\xspace}}
\tsc{WGM}
\tsc{QE}
\tsc{EP}
\tsc{PMS}
\tsc{BEC}
\tsc{DE}

\newcommand{\logt}[1]{\ensuremath{\log_{10}\left(#1\right)}}

\newcommand{\lw}{\ensuremath{l^\mathrm{W}}}
\newcommand{\lwx}[1]{\ensuremath{l^{\mathrm{W,}#1}}}
\newcommand{\lp}{\ensuremath{l^\mathrm{p}}}

\newcommand{\pml}{\ensuremath{~\substack{\leq + \\ \geq -} ~}}
\newcommand{\Pel}{\ensuremath{po}}

\begin{document}
\let\WriteBookmarks\relax
\def\floatpagepagefraction{1}
\def\textpagefraction{.001}
\shorttitle{Ventilation System Optimization: Airflow and Acoustics}
\shortauthors{JHP Breuer et~al.}


\title [mode = title]{Algorithmic Planning of Ventilation Systems: Optimising for Life-Cycle Costs and Acoustic Comfort}




\author[1]{Julius H. P. Breuer}[type=editor,
                        auid=000,bioid=1,
                        orcid=0000-0002-6226-7208]

\ead{julius.breuer@fst.tu-darmstadt.de}
\credit{Conceptualisation, Methodology, Software, Validation, Formal analysis, Investigation, Resources, Visualisation, Writing - First Draft, Writing - Review and Editing}

\author[1]{Peter F. Pelz}
\ead{peter.pelz@fst.tu-darmstadt.de}
\credit{Funding Acquisition, Supervision, Project Administration}
\cormark[1]

\address[1]{Chair of Fluid Systems, TU Darmstadt, Otto-Berndt-Straße 2, 64287 Darmstadt, Germany}
\cortext[cor1]{Corresponding author}

\begin{abstract}
The European Union's climate targets challenge the building sector to reduce energy use while ensuring comfort.
Ventilation systems play an important role in achieving these goals. During system planning, the primary focus tends to lie on reducing life-cycle costs, including energy and investment expenses. Acoustic considerations which contribute significantly to occupant comfort, are either addressed as an afterthought or overlooked. This can result in suboptimal designs, where silencers are added indiscriminately without properly assessing their necessity.
This paper introduces a novel method for optimising life-cycle costs through mathematical optimisation while adhering to predefined noise limits. We propose new model equations with reduce non-linearity better suited for integration into the optimisation framework. Further, they present a comprehensive approach to optimising ventilation systems under multiple load scenarios. Our method surpasses the traditional sequential approach by enabling simultaneous consideration of airflow and acoustics in a single, holistic optimisation step.
A case study demonstrates the method’s practical application, showing that optimal solutions can be computed efficiently. The results reveal that, with appropriate fan selection, many silencers can be eliminated. Additionally, the method supports decision-making by transparently illustrating the trade-offs between life-cycle costs and noise limits. Notably, while optimal solutions from the sequential and holistic approaches align for most noise limits, the holistic method achieves a 12~\% reduction in costs under specific noise constraints.
These results demonstrate the benefits of integrating airflow and acoustic design while underscoring the need for further application on more diverse building types and more complex ventilation system configurations.
\end{abstract}

\begin{highlights}
\item Mastering the complexity of acoustic phenomena in ventilation system design
\item Coupling airflow and acoustics in a Mixed-Integer Nonlinear Program
\item Deriving new model equations for fans, volume flow controllers and silencers
\item Trade-off between life-cycle costs and noise limits enables transparent decisions.
\item Holistic optimisation allows life-cycle costs to be reduced by up to 12~\%
\end{highlights}


\maketitle

\input{introduction}

\input{Methods}

\input{component_models}

\input{Results}

\input{Discussion}

\appendix

\input{Appendix}

\newpage

\printcredits

\section*{Competing interests}
There is NO competing interest.

\section*{Acknowledgement}
The presented results were obtained within the research project “Algorithmic System Planning of Air Handling Units”, Project No. 22289 N/1, funded by the program for promoting the Industrial Collective Research (IGF) of the German Ministry of Economic Affairs and Climate Action (BMWK), approved by the Deutsches Zentrum für Luft- und Raumfahrt (DLR). We want to thank all the participants of the working group for the constructive collaboration.

\section*{Generative AI and AI-assisted technologies in the writing process}
During the preparation of this work the authors used GPT-4 in order to assist in revising the style and use of language. After using this tool/service, the authors reviewed and edited the content as needed and take full responsibility for the content of the publication.

\bibliographystyle{elsarticle-num}

\bibliography{main}


\end{document}

%% file: introduction.tex

\section{Introduction}
\label{sec:intro}
EU climate targets impose significant requirements on the building sector, driving the need for more efficient building operation \cite{FederalMinistryforEconomicAffairsandEnergy.2015}. To reduce heat losses, buildings become better insulated and thus more airtight, requiring mechanical ventilation systems to assure good indoor air quality while maintaining energy-efficient operation \cite{carrie2017impact}.

Ventilation systems are, however, a major source of noise in buildings. While their primary goal is ensuring good indoor air quality, they should not compromise the acoustic comfort of the building occupants~\cite{en12081414}. This is particularly important as they are considered a major source of noise. This can be summarised as two conflicting objectives: (i) low life-cycle costs and (ii) low noise present a notable challenge, as quieter systems typically require additional components, which increases the costs and energy consumption.
\newline
In this paper, we propose a novel optimisation framework to address these conflicting objectives. Our approach integrates airflow and acoustic considerations into a unified optimisation process, enabling a holistic design of ventilation systems that minimises life-cycle costs while adhering to predefined noise limits. This method overcomes the limitations of traditional sequential planning by considering both airflow and noise factors simultaneously.
\newline
Traditional planning procedures often fall short in resolving this conflict. During the detailed planning phase, component selection and placement are carried out (see \cref{app:planning_procedures} for a comprehensive overview). In this phase, the system is developed in a stepwise manner: initially, components required to meet airflow demands are designed with a focus on minimizing costs and energy consumption. Noise limits are subsequently addressed by adding silencers or, if necessary, iteratively refining the prior design. However, components such as silencers introduce trade-offs, including increased pressure loss and flow noise, which negatively affect energy efficiency and overall costs. This underscores the complexity of managing such a multi-physics, techno-economic planning challenge.
Guidelines and standards provide simple design rules for efficient system operation, such as demand-responsive systems and component efficiency recommendations~\cite{VDI3803-1,en16041853,DeutscheEnergieAgentur.2018}. These rules, based on approximate calculations, often fail to capture the full range of potential solutions. While life-cycle cost analysis is recommended to balance investment costs and energy efficiency \cite{stanford2019analysis}, such methods are typically limited to comparing only a few predefined variants. Consequently, the resulting designs deviate from the optimal solution, leaving significant performance potential unrealised.
An equally critical aspect is the correct dimensioning and selection of fans, factoring in load profiles. Fan selection should not be based solely on efficiency in the maximum load scenario but should instead focus on optimizing annual energy costs by considering the characteristic diagram, expected load profile, fan type, size, and speed \cite{DeutscheEnergieAgenturGmbH.2010,mckane2003improving}. However, standards like EN 16798–3~\cite{DIN16798-3} overlook this complexity, as they calculate energy consumption based solely on a single design load point, neglecting the broader operational variability.
\newline\newline
These limitations emphasise the growing demand for algorithmic planning procedures, a trend increasingly reflected in recent research. Broadly, algorithmic design can be categorised into two approaches: data-driven heuristic methods and physics-based methods utilising mathematical optimisation. For these methodologies to be effective, a formalised approach to the planning task is essential. Reviews by Ala'raj et al., Sha et al., and Ahmad et al. provide valuable insights into data-driven heuristic methods, which largely focus on the operational aspects of ventilation systems~\cite{Alaraj.2022,sha2019overview,ahmad2016computational}. However, when it comes to design considerations, aspects like fan selection and placement are often neglected, with the focus typically placed on the ventilation system as a whole but only at a low-granularity level~\cite{sha2019overview}. As a result, critical elements such as the exploration and evaluation of more efficient system variants remain largely unaddressed, further highlighting the need for more comprehensive and systematic planning approaches.
As data-driven methods are a type of heuristic, they do not guarantee optimal or near-optimal results. In contrast, mathematical optimisation methods can identify globally optimal solutions. These methods enable the incorporation of problem-specific knowledge, which is invaluable for system-wide optimisations where interactions between components significantly influence both costs and performance. In the context of ventilation, many studies focus on mine ventilation systems~\cite{acuna2014review}. However, recent research by Schänzle et al.~\cite{schanzle2015good} and Müller et al.~\cite{muller2023planning} has demonstrated the effectiveness of mathematical optimisation in minimising life-cycle costs of a building's ventilation system. Notably, Müller et~al. achieved reductions not only by incorporating multiple load cases but through more efficient system topologies, such as distributed fan setups. These optimisations led to a $22~\%$ reduction in life-cycle costs, including a $28~\%$ decrease in energy costs. Their approach models the planning procedure as a Mixed-Integer Nonlinear Program (MINLP), which accounts for the nonlinear characteristic curves of fans and integrates both binary and continuous decision variables within the planning and operation process.\newline
Despite advancements in related fields, a significant research gap persists regarding the integration of acoustics in the design of ventilation systems for building applications \cite{Alaraj.2022}. While standards exist to calculate noise production, they do not provide guidelines for incorporating noise control as a criterion for optimal system design. Consequently, acoustics are often addressed retroactively, which can undermine the low life-cycle cost results achieved earlier.
In contrast, fields such as automotive HVAC systems have made significant strides in acoustics optimisation with high level of detail~\cite{eilemann1999practical,luzzato2019aero}. However, this approach is not applicable to building ventilation systems due to their larger scale and complexity. Ferrara et al. explored acoustics optimisation using genetic algorithms, though their work concentrated on the building envelope rather than internal ventilation systems \cite{Ferrara.2021}. Similarly, Ala'ray et al. identified a lack of data-driven methods, and, to the best of our knowledge, no optimisation approaches currently incorporate acoustics into the planning of ventilation systems \cite{Alaraj.2022}.
This oversight leads to potentially inefficient designs, costly decision-making processes, and limited transparency during planning. Human planners, who must navigate a vast array of components and configurations, often face difficulties balancing energy efficiency, noise reduction, and cost-effectiveness~\cite{muller2023planning}. 
\newline\newline
In the paper, we address the above mentioned issues by formulating a physics-based optimisation methodology that, for the first time, couples acoustics with life-cycle cost optimisation in ventilation systems. To address these issues in the scope of this paper, we define the research questions as follows:
\begin{enumerate}
    \item Can the design of ventilation systems including acoustics using mathematical optimisation be automated and solved within an acceptable timeframe?
    \item Can the proposed methodology be used to make the conflict between life-cycle costs (including energy costs) and acoustic comfort transparent for decision makers?
    \item Does the proposed methodology allow for a comprehensive comparison between holistically planned and sequentially planned ventilation systems?
\end{enumerate}

Our aim is to bridge the existing research gap by developing an approach that integrates acoustic considerations with energy efficiency and economic goals. This approach acknowledges that acoustic requirements are not merely regulatory hurdles but integral design constraints that could influence the overall efficiency of ventilation systems. The resulting methodology aims at making the connection between life-cycle costs and acoustics transparent for decision makers. It further allows a holistic approach, where airflow and acoustics are considered and all components are laid out in one step. This is compared to the traditional sequential approach of first laying out the airflow and secondly the acoustics.
Therefore, a physical representation of the ventilation system ensuring airflow is introduced in \cref{sec:airflow_modelling}. Then, additions to the model are introduced in \cref{sec:acoustic_modeling} coping with the challenges of acoustics. Solving the optimisation problem yields globally-optimal solutions that allow for objective decision-making in terms of conflicting goals as detailed in  \cref{sec:solving_algorithm}. The ventilation system's component models for both airflow and acoustics are introduced in \cref{sec:component_models}. Using methods from mathematical optimisation ensures that each problem step is solved with minimal possible life-cycle costs. In a case study, the method is applied to a real-world example (\cref{sec:results}) analysing its viability, making the optimal decision-space transparent and comparing it to the traditional planning procedures.

%% file: Methods.tex
\section{Methods}

The design task is first described in greater detail. Then, airflow modeling and the acoustic modeling are presented. Finally, a problem-specific solving algorithm is introduced.

\subsection{Overview}
\label{sec:method_overview}

\begin{figure*}[htb]
	\centering
		\includegraphics[width=0.9\linewidth]{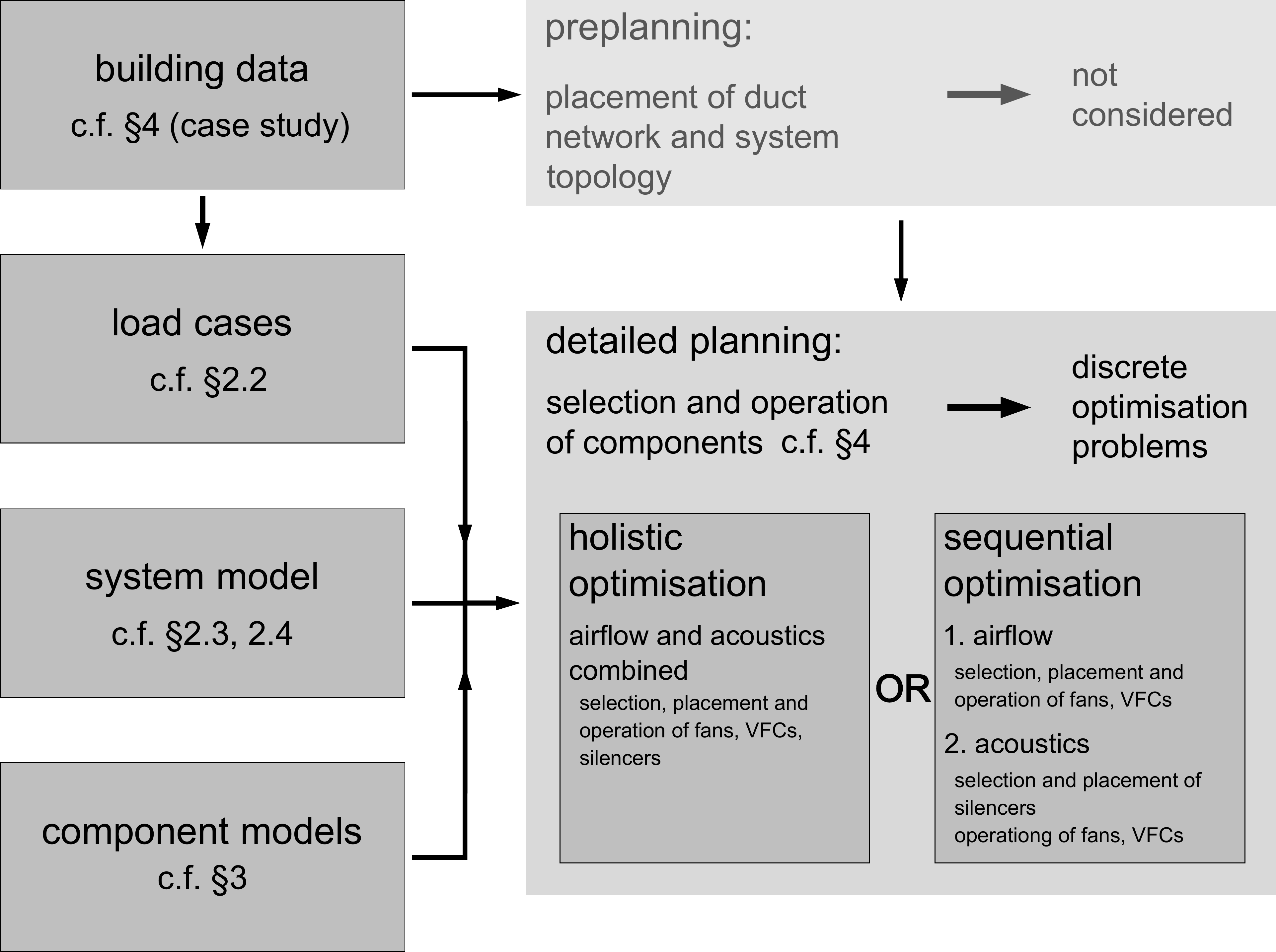}
	\caption{procedure of the planning approach and organisation of the manuscript.}
	\label{fig:method_overview}
\end{figure*}

The aim of the presented method of optimisation ventilation systems is to allow transparent and efficient decision-making. The procedure used in this manuscript including all input data is described in more detail in \cref{fig:method_overview}. 
In many countries, regulatory frameworks define the planning process. Despite varying in rigor, detail, and scope, these regulations generally encompass three main phases: preplanning, detailed planning, and construction oversight. See \cref{app:planning_procedures} for a comparison between multiple western countries. The method provided here focuses on the detailed planning and is in accordance with the real planning processes as outlined in \cref{app:planning_procedures}. Starting from building data, see \cref{sec:case_study}, load profiles accounting for outdoor air demand are obtained for the individual rooms, see \cref{sec:load_cases}. Based on these, the duct network and the feasible component locations are considered given. Allowing to account for different ventilation system topologies and to build upon expert knowledge in the early stages of the planning process, this step is considered as already done in this manuscript. Using load profiles, system and components models as well as the pre-planned solution, the detailed planning can be performed. Its result has to be finalised by human planners.

The procedure ensures that the optimisation model focuses on the most relevant aspects of the detailed planning and thus is both practical and relevant to the real-world demands of ventilation system design.

\subsection{Load Cases}
\label{sec:load_cases}

Ventilation systems are planned to fulfill outdoor air demands. The usual practice is designing solely for the maximum load, i.e. the maximally required volume flow in all rooms. To depart from this practice, varying demands are accomodated, and to achieve this, a range of load cases is examined, each characterised by different air flows and associated pressure losses.  The subsequent section details how to establish load profiles for different rooms in a building.

To establish the load profiles, it is essential to determine the necessary airflow. Three different types of airflow calculations are utilised: the building-specific airflow ($Q^\mathrm{build}$), the per-person airflow ($Q^{\mathrm{person}}$), and a case-specific design airflow ($Q^\mathrm{plan}$). The building-specific airflow, calculated according to DIN V 18599–10~\cite{DIN.18599-10}, meets the outdoor air needs for each room and varies depending on room type and size. The European standard EN 16798–1 offers detailed hourly occupancy profiles for various room types~\cite{DIN16798}. For rooms not covered in the standard, occupancy profiles are developed based on the author's discretion and under the guidelines of DIN V 18599–10. Combining these profiles with the per-person airflow rate, $Q^{\mathrm{person}}$, allows for the creation of a load profile dependent on occupancy. Typically, these two airflow types suffice. However, for the case study presented later, the actual design airflow $Q^\mathrm{plan}$, determined by a technical planning expert, is also considered. This value serves as a benchmark in evaluating the determined airflow rates. Based on these volume flows, the required volume flow for a certain room and a certain scenario $Q^{\mathrm{req}}$ corresponding to Alsen is determined as the maximum of the three flows~\cite{Alsen.2016}, as shown in \cref{eq:Qreq}. Here, $\tilde{a}$ indicates the percentage of occupancy, and $a^{\mathrm{max}}$ is the maximum number of people in the room.

\begin{equation} \label{eq:Qreq}
Q^{\mathrm{req}} = \max \left\{ Q^{\mathrm{build}}, \; \tilde{a} \cdot a^{\mathrm{max}} \cdot Q^{\mathrm{person}}, \; \tilde{a} \cdot Q^{\mathrm{plan}} \right\}
\end{equation}

The selection of the maximum ensures that at least the volume flow required due to the room size or the occupancy is always conveyed.

According to European standard EN 16798–1, rooms are typically ventilated for a maximum of 14 hours per day. Consequently, this methodology leads to the identification of 14 distinct load cases. In order to reduce the amount of load cases, clustering is employed using the k-means algorithm~\cite{MacQueen.1967}. This process involves treating the volume flow requirements of each room for a specific operating hour as an individual data point. Consequently, the 14 data points are clustered together. A notable aspect of the k-means algorithm is its requirement for a predefined number of clusters, which is advantageous for generating a set number of clusters.

After the establishment of load cases, the next step is to determine the pressure loss attributed to the ductwork. The most prevalent method for this involves the utilisation of CAD design software, integrated with a calculation of pressure loss that considers the various components of the ductwork. This approach needs to be performed for each load case.

\subsection{Airflow modelling}
\label{sec:airflow_modelling}

\begin{figure}[htbp]
	\centering
		\includegraphics[width=0.8\linewidth]{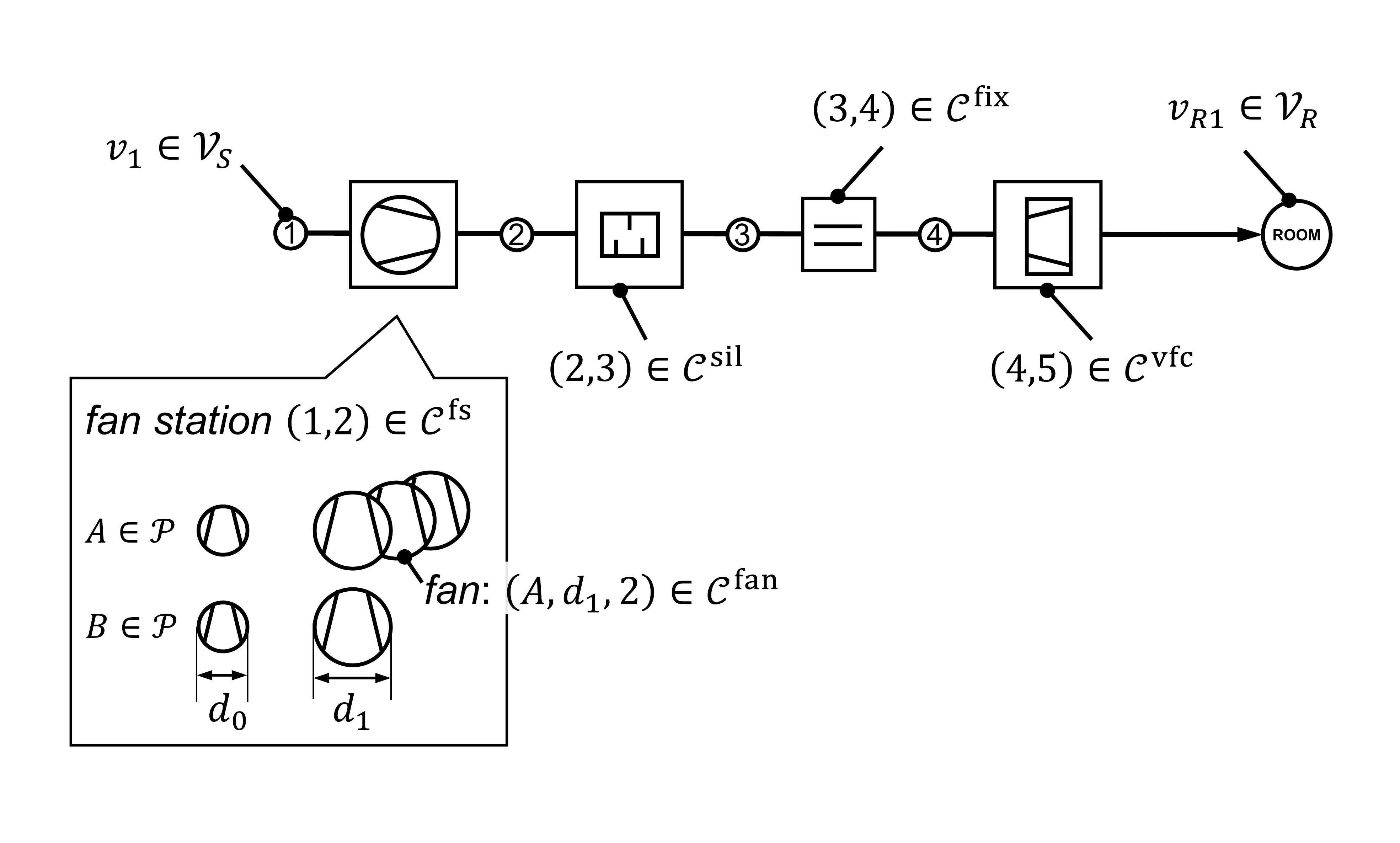}
	\caption{example configuration as a graph representation of a simple ventilation system with different components on the edges. }
	\label{fig:component_sets_example}
\end{figure}

As supply and exhaust air system are modelled separately, the ventilation system can be defined as a mathematical tree-like graph $\mathcal{G}(\mathcal{V},\mathcal{E})$, with nodes $v\in\mathcal{V}$ and edges $(i,j)\in\mathcal{E}$. The ventilation system model is introduced as a Mixed-Integer Nonlinear Program (MINLP). The model contains the objective, the mathematical graph of the system, the possible location of variable components as well as the locations of fixed components. Moreover, the load cases $\mathcal{S}$, topology constraints and conservation equations are defined in the system model. The mathematical optimisation problem is structured in two stages: Firstly, identifying a cost-effective investment decision for the components. Secondly, operating a subset of the selected components for each load case within the specified load profile, ensuring that the system meets the demand while maximising efficiency. This makes the problem a 2-stage stochastic MINLP. The ventilation system's sets are given in \cref{tab:set} and the parameters and variables are given in \cref{tab:param_var_sys}. An example setup is shown in \cref{fig:component_sets_example}.
In order to define the required volume flows in each load case, rooms are given as target nodes. To fulfill these, component models are defined on the graphs edges. The components, namely fan stations, volume flow controllers (VFCs) and silencers are placed on the edges where they cause a change in pressure and octave sound power level. Fans are not directly placed on the edges but in fan stations, where multiple fans can occur in parallel. Component models are introduced in detail in \cref{sec:component_models}.

To ensure clarity, optimization variables are written in lowercase, while parameters are written in uppercase letters.

The objective of the optimisation is to minimise life-cycle costs. These include energy, investment and maintenance costs. The individual costs are calculated in accordance to the annuity method as defined in VDI 2067. Here, maintenance (M) costs and the service (S) costs are proportional to the investment costs, simplifying the calculation. In the following the sum of investment, maintenance and service costs is referred to as investment costs. The interest rate $Z$ is factored into the calculations. The price change factor for electricity (E) $R^\mathrm{E}$ is derived from averaging the electricity prices $C^\mathrm{E}$ over the last 20 years. The maintenance and service price change $R^\mathrm{M,S}$ is taken from VDI 2067 \cite{VDI2067}.
The price-dynamic present values $B^\mathrm{E}$ and $B^\mathrm{M,S}$ can be calculated using
\begin{align}
    B^m = \frac{1-R^m/Z}{Z-R^m} \qquad \forall m \in \{\mathrm{M,S},~\mathrm{E} \}.
\end{align}
The annuity factor $A$ is calculated using the relation
\begin{align}
    A = \frac{Z - 1}{1 - Z ^{-T^\mathrm{use}}}.
\end{align}
\newline
The objective is given as the sum of investment costs and operation costs. The total investment costs include costs $c^{com}$ for fan stations (including fan costs), silencers and VFCs. The total operation costs include each fan station's operation costs due to power consumption $\Pel$ in each load case $s\in\mathcal{S}$:
\begin{equation}
    \min \, \, T^\mathrm{use}\sum_{com} \sum_{(i,j)\in\mathcal{C}^{com}} A^{\mathrm{F}, com} c^{com}_{i,j} + C^\mathrm{E}AB^\mathrm{E}T^\mathrm{use} \sum_{s\in\mathcal{S}} T_s \sum_{(i,j)\in\mathcal{C}^\mathrm{fs}} \Pel_{i,j,s},
\end{equation}
with $A^{\mathrm{F}, com}$ being the resulting annuity investment cost factors per component ($com\in \{\text{fs, sil, vfc}\}$). Their calculation is given in \cref{app:annuity}. $T^\mathrm{use}$ being the entire operating time of the ventilation system in hours and $T_s$ being the relative frequency of the respective load case.

Every component shares a connection to the ventilation system model through input and output variables, denoted by index '$\mathrm{in}$' and '$\mathrm{out}$'. Furthermore, every component can only be activated if it is purchased. The source and target pressures as well as the connections between system and component models are defined with the constraints \cref{eq:energy_conservation1,eq:energy_conservation2,eq:energy_conservation3,eq:energy_conservation4}. To ensure continuity, the inflow at each node has to equal its outflow, \cref{eq:continuity}. As the volume flows on the edges can be calculated a-priori, $Q_{s,i,j}$ is a parameter and continuity, \cref{eq:continuity}, holds implicitly.
\begin{align}
    \label{eq:energy_conservation1}
    p_{s,i} &= p^{\mathrm{S}}_{s,i} \qquad &\forall s\in\mathcal{S} \quad \forall i \in \mathcal{V}_\mathrm{S} \\
    \label{eq:energy_conservation2} p_{s,i} &= p^{\mathrm{T}}_{s,i} \qquad &\forall s\in\mathcal{S} \quad \forall i \in \mathcal{V}_\mathrm{T} \\
    \label{eq:energy_conservation3} p_{s,i} &= p^{\mathrm{in}}_{s,i,j} \qquad &\forall s\in\mathcal{S} \quad \forall (i,j) \in \mathcal{C}\\
    \label{eq:energy_conservation4} p_{s,j} &= p^{\mathrm{out}}_{s,i,j} \qquad &\forall s\in\mathcal{S} \quad \forall (i,j) \in \mathcal{C}
    \\
    \label{eq:continuity}
    \sum_{i\in \mathcal{V}} q_{s,i,j} &= \sum_{k\in \mathcal{V}} q_{s,j,k} \qquad &\forall s \in \mathcal{S} \quad \forall j\in \mathcal{V} / \mathcal{V}_\mathrm{S} / \mathcal{V}_\mathrm{T}.
\end{align}

\begin{table}[htbp]
\centering
\caption{Used parameters and variables in the system.}
\label{tab:param_var_sys}
\begin{tabular}{  p{2cm} p{4cm} >{\raggedright}p{8cm} }
 parameter & domain  & description \\ 
 \hline
 $A$ & $\mathbb{R}^+$ & annuity factor\\
 $A^{\mathrm{F},com}$ & $\mathbb{R}^+$ & annuity investment cost factor for component $com\in \{\text{fs, sil, vfc}\}$\\
 $B^\mathrm{S,M}$, $B^\mathrm{E}$ & 1.03, 1.08 & price-dynamic present value \\
 $F^{\mathrm{S},com}$, $F^{\mathrm{M},com}$ & $\mathbb{R}^+$, $\mathbb{R}^+$ & cost factor of maintenance and service for component $com\in \{\text{fs, sil, vfc}\}$. Used in \cref{app:annuity}\\
 $C^\mathrm{E}$ & $\SI{0.15}{\perkWh}$) & electricity price \\
 $T^\mathrm{use}$ & $\mathbb{R}^+$ & time of usage in hours for all years of operation. In accordance with VDI 2067, 12 years $\times$ 250 days/year $\times$ 14 hours/day is used.\\
 $T^{\mathrm{dep},com}$ & $\mathbb{R}^+$ & time in years until component $com\in \{\text{fs, sil, vfc}\}$ is deprecated. Used in \cref{app:annuity}\\
 $T_s$ & [0,1] & relative frequency of load case $s\in\mathcal{S}$\\
 $Z$ & 1.07 & interest rate \\ 
 $\overline{\eta}^\mathrm{fan}$ & $[0,1]$ & maximal fan efficiency in the ventilation system\\
 $p_{i,s}^\mathrm{S}$ & $\mathbb{R}^+$ & pressure at source $i\in\mathcal{V}\mathrm{S}$ in scenario $s\in\mathcal{S}$ \\
 $p_{i,s}^\mathrm{T}$ & $\mathbb{R}^+$  &pressure at target $i\in\mathcal{V}\mathrm{T}$ in scenario $s\in\mathcal{S}$\\
 $\overline{\Delta p}, \overline{\lw}$ & $\mathbb{R}^+$ & upper limits used for bigM constraints.
 \\
 \\
 variable & range  & description \\
 \hline
 $p_{i,j,s}^\mathrm{in}$ & $[0,\overline{\Delta p}]$ & input pressure into component $(i,j)\in\mathcal{C}$ in load case $s\in\mathcal{S}$\\
 $p_{i,j,s}^\mathrm{out}$ & $[0,\overline{\Delta p}]$ & output pressure from component $(i,j)\in\mathcal{C}$ in load case $s\in\mathcal{S}$\\
 
\end{tabular}
\end{table}

\begin{table}[htbp]
\centering
\caption{Sets used for airflow and acoustic optimisation problems.}
\label{tab:set}
\begin{tabular}{  p{3cm} p{7cm} }
  set  & description \\ 
  \hline
  $\mathcal{F}$ & set of octave bands\\
 $\mathcal{S}$ & set of load cases \\
 $\mathcal{E}$ & set of edges\\
 $\mathcal{C}^\mathrm{fs}$ & set of fan stations\\
 $\mathcal{C}^\mathrm{sil}$ & set of silencers\\
 $\mathcal{C}^\mathrm{vfc}$ & set of VFCs\\
 $\mathcal{C}^\mathrm{fix}$ & set of fixed components\\
 $\mathcal{C}$ & set of all components $\mathcal{C}=(\mathcal{C}^\mathrm{fs}\times\mathcal{C}^\mathrm{sil}\times\mathcal{C}^\mathrm{vfc}\times\mathcal{C}^\mathrm{fix})$ \\
 $\mathcal{P}$ & set of fan product lines\\
 $\mathcal{D}$ & set of fan diameters\\
 $\mathcal{N}$ & set used for numbering identical fans.\\
 $\mathcal{C}^\mathrm{fan}$ & set of fans $\mathcal{C}^\mathrm{fan} \subseteq (\mathcal{P}\times\mathcal{D}\times\mathcal{N})$\\
 $\mathcal{V}$ & set of vertices\\
 $\mathcal{V}_\mathrm{S}$ & set of sources where a pressure is predefined\\
 $\mathcal{V}_\mathrm{R}$ & set of rooms\\
 $\mathcal{V}_\mathrm{T}$ & set of targets where a pressure is predefined\\
 
\end{tabular}
\end{table}

\subsection{Acoustic modeling}
\label{sec:acoustic_modeling}

In this subsection, the airflow-only model is extended for introducing the coupled airflow and acoustics model. This separation is possible because the airflow model does not depend on the acoustic equations, thus the coupling is one sided. To derive the model, first acoustic modeling approaches are discussed, then room models and a method for simplifying the level addition from Breuer et~al.\cite{Breuer2024} are introduced.
Finally, an iterative solving algorithm is introduced to obtain a Pareto-front for the conflicting goals minimal life-cycle costs and minimal noise limits.

In buildings, limitations of noise levels are defined for various rooms. These noise limits are usually set by the building owner according to standards or their own specifications in form of sound pressure levels. Sound pressure levels $\lp$ are the pressure fluctuations that a person perceives at a certain point in the room. They are defined as the logarithm of the squared sound pressure $p$ and can be formulated as:
\begin{equation}
    \lp = 20 \logt{\frac{p}{p^\mathrm{ref}}},
\end{equation}
with $p^\mathrm{ref} = \SI{2e-5}{\pascal}$ being the reference sound pressure~\cite{VDI2081}. As the human ear perceives sound pressure differently across frequencies, it is common practice to apply a weighting to different frequency levels. The A-weighted sound pressure level $l^{p,\mathrm{A}}$ is obtained by adding the parameter $l^\mathrm{A}$, see \cref{tab:a_weight}, according to the sound pressure level's frequency~$f$:
\begin{equation}
    l^\mathrm{p,A} = \lp + l^\mathrm{A}.
\end{equation}

\begin{table}[htbp]
        \centering
      \captionof{table}{A-weighting parameters for octave frequency bands $\mathcal{F}$}
      \begin{tabular}[t]{l | r}
       $f$ in $\si{\hertz}$ & $l^{\mathrm{A}}$ in $\si{\decibel}$\\
      \hline
      45 - 88 & -25.2\\
      88 - 177 &-15.6 \\
      177 - 354 & -8.4 \\
      354 - 707 & -3.1\\
      707 - 1414 & 0\\
      1414 - 2828 & 1.2\\
      2828 - 5657 & 0.9\\
      5657 - 11314 & -1.1\\
      \end{tabular}
      \label{tab:a_weight}
  \end{table}

To ensure that noise limits are not exceeded, embedding acoustic modelling in the design process is necessary. As ventilation systems provide a main source of noise in buildings, a model connecting the noise generated by the ventilation system with the noise limits is created. However, the physically correct modeling of sound waves is a very complex task that requires detailed numerical simulations and cannot be achieved for entire ventilation systems. To address this, common methodologies in noise modeling simplify the propagation of sound waves into a zero-dimensional (0D) phenomenon. By adopting this simplification, a system model can be formulated, consisting of various individual elements and their interconnections.

Among these methodologies, two primary approaches are predominant. These approaches are valid for different Helmholtz numbers:
\begin{align}
    \mathrm{He} = \frac{d_\mathrm{char}}{\lambda} = \frac{d_\mathrm{char} f}{c},
\end{align}
with $d_\mathrm{char}$ being a characteristic length, i.e. the duct diameter, $\lambda$ being the wave length, $f$ the respective frequency and $c$ the speed of sound.

\begin{figure*}[htpb]
     \centering
     \begin{subfigure}[T]{0.45\textwidth}
        \centering
    	\includegraphics[width=\textwidth]{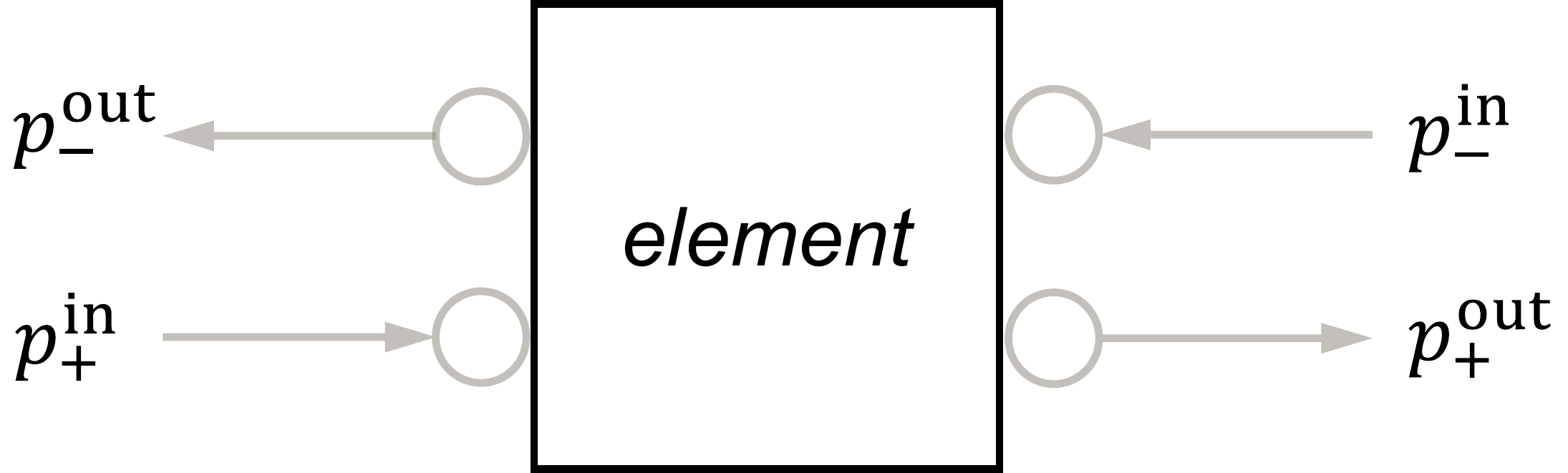}
    	\caption{pressure fluctuations 2-pole, representing a left and a right wave of pressure fluctuations}
    	\label{fig:2port}
     \end{subfigure}
     \hfill
     \begin{subfigure}[T]{0.5\textwidth}
         \centering
         \includegraphics[width=\textwidth]{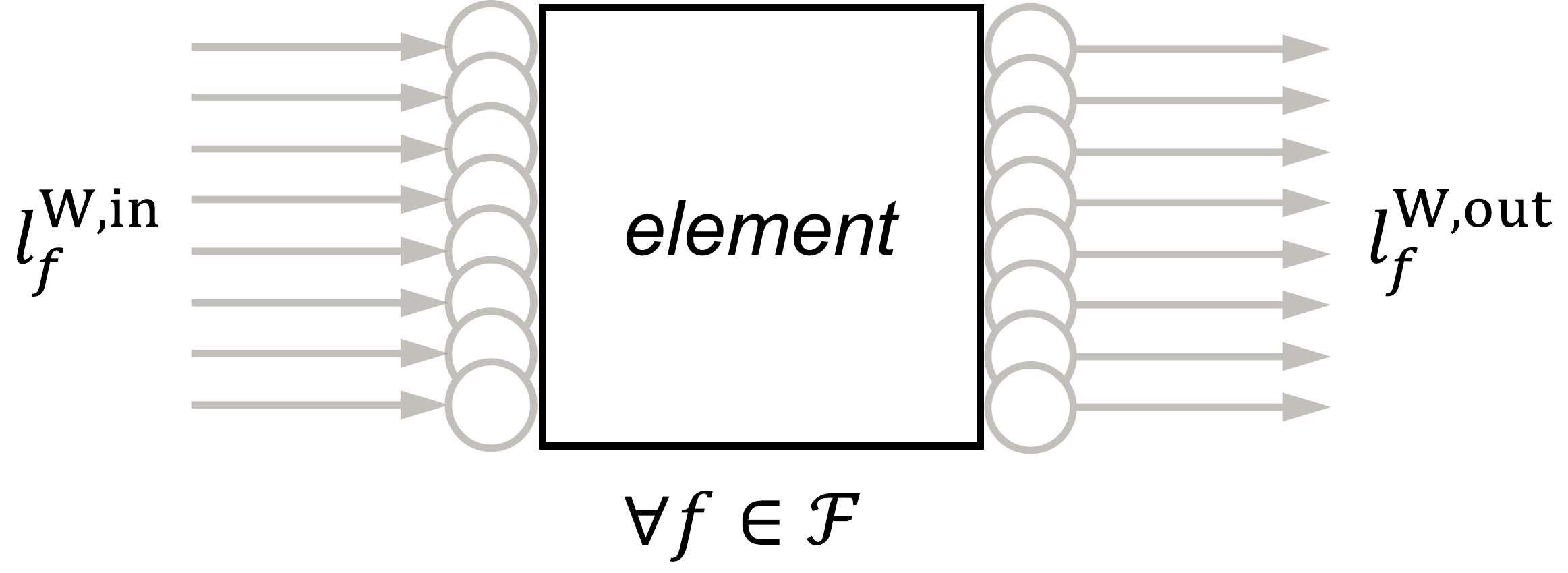}
         \caption{8-pole source-path-receiver model, one pole for each frequency band.}
         \label{fig:8port}
     \end{subfigure}
     \caption{two different acoustic 0D model approaches}
    \label{fig:port_approaches}
\end{figure*}

The first approach is valid for lower Helmholtz numbers $\mathrm{He}\ll 1$ where diffraction of waves plays an important role. The approach is normally used in the low frequency range \cite{WagihNashed.2018}. It stems from the wave equation in acoustics and focuses on the amplitudes of pressure fluctuations and, in some cases, velocity fluctuations, of forward and backward waves~\cite{glav.1997}. This method allows for the examination of changes in these variables across each component of the ventilation system. Consequently, the system's elements are categorised as 2-poles or 4-poles (if velocity fluctuations are included), reflecting the number of variables considered in the forward and backward directions. An example element for the approach is shown in \cref{fig:2port}.

The other approach uses a one directional model, the source-path-receiver model. Here, the acoustic property that flows through the ventilation system is the sound power level \lw. While the 2-port is mostly used in pipe systems, the source-path-receiver model is mostly used in ventilation systems due its viability at larger duct diameters, i.e. at high Helmholtz numbers $\mathrm{He}\geq 1$ and high-frequency ranges \cite{WagihNashed.2018}.
The sound power level is defined as:
\begin{equation}
    \lw = 10\logt{\frac{po}{po_0}}.
\end{equation}

As levels often have to be added up for further calculation, the mathematical operation of level addition is defined as standard summation is invalid. For adding levels $\lw_1$ and $\lw_2$, level addition is defined as follows:
\begin{align}
    \label{eq:level_add}
    \lw_3 = 10\logt{10^{0.1\lw_1}+10^{0.1\lw_2}} = \lw_1 ~\tilde{+}~ \lw_2,
\end{align}
with $\tilde{+}$ introduced as the level addition symbol.

The sound power level change of a component is strongly linked to the frequency. As such, it is reasonable to calculate the sound power level in eight octave bands, yielding an 8-pole with eight octave sound power levels $\lw_f$~\cite{VDI2081} with the frequency bands $f\in\mathcal{F}$. To denote that an octave sound power/pressure level is meant, a subscript $f$ is used for the respective frequency. Inside each 8-pole component, the octave sound power level changes for two reasons, acoustic dampening of the component $d^\mathrm{W}_f$ and additional flow noise $\lwx{\mathrm{flow}}_f$. The input octave sound power level is first damped and then level added to the octave sound power level of the flow noise:
\begin{align}
    \label{eq:}
    \lwx{\mathrm{out}}_f = \left(\lwx{\mathrm{in}}_f-d^\mathrm{W}_f \right) ~\tilde{+}~ \lwx{\mathrm{flow}}_f.
\end{align}

Besides these two approaches, mixed-models are proposed that contain source-path-receiver models for high and 2-poles~\cite{gijrath2002} for low frequencies.

The source-path-receiver approach is prescribed for ventilation system planning in the form of the ASHRAE \emph{Algorithms for HVAC Acoustics}~\cite{ReynoldsBledsoe1990} and VDI standard 2081~\cite{VDI2081}. A disadvantage of the source-path-receiver model is that it does not consider effects that are particularly important at low frequencies, like standing wave effects which also change the power output~\cite{gijrath2002}.
Besides this disadvantage, this manuscript adopts the source-path-receiver model in accordance with VDI 2081. The main reason is that the model has two significant advantages compared to the 2-pole or 4-pole: It is compatible with readily available data from manufacturers and its application is the mandatory standard to ensure proper indoor noise levels. Thus, it presents an industry-relevant and robust method.

\begin{figure}[htb]
	\centering
		\includegraphics[width=0.5\linewidth]{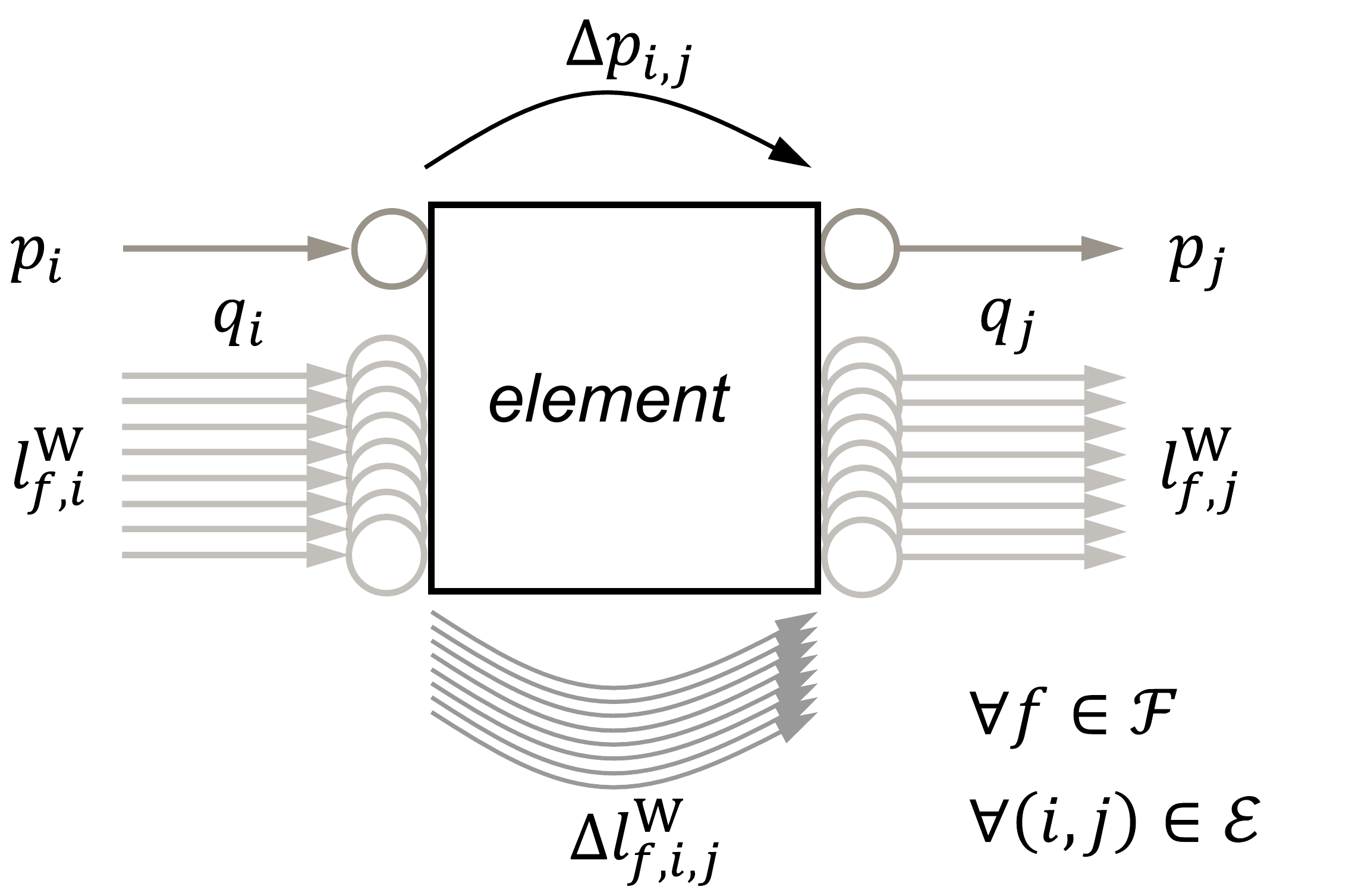}
	\caption{generic 9-pole element. Figure adapted from Breuer et al.~\cite{Breuer2024}}
	\label{fig:multipole_element}
\end{figure}

As airborne sound propagation is the main source of noise in the rooms, a detailed model of the ventilation system with regard to acoustics has to be combined with pressure and volume flow modelling. Therefore, the 9-pole approach according to Breuer et~al.~\cite{Breuer2024} is used, see \cref{fig:multipole_element}. The 9-pole describes elements by their change in pressure and octave sound power levels.

For the sake of brevity, the term flow noise is used to refer to \emph{octave sound power level of the flow noise} and \emph{sound dampening} refers to the \emph{octave sound power level of dampening}. Furthermore, sound pressure level limits are referred to as noise limits.

\subsubsection{Acoustic room models}
\label{sec:acoustic_room_models}
\begin{figure*}[htbp]
	\centering
	\includegraphics[width=0.7\linewidth]{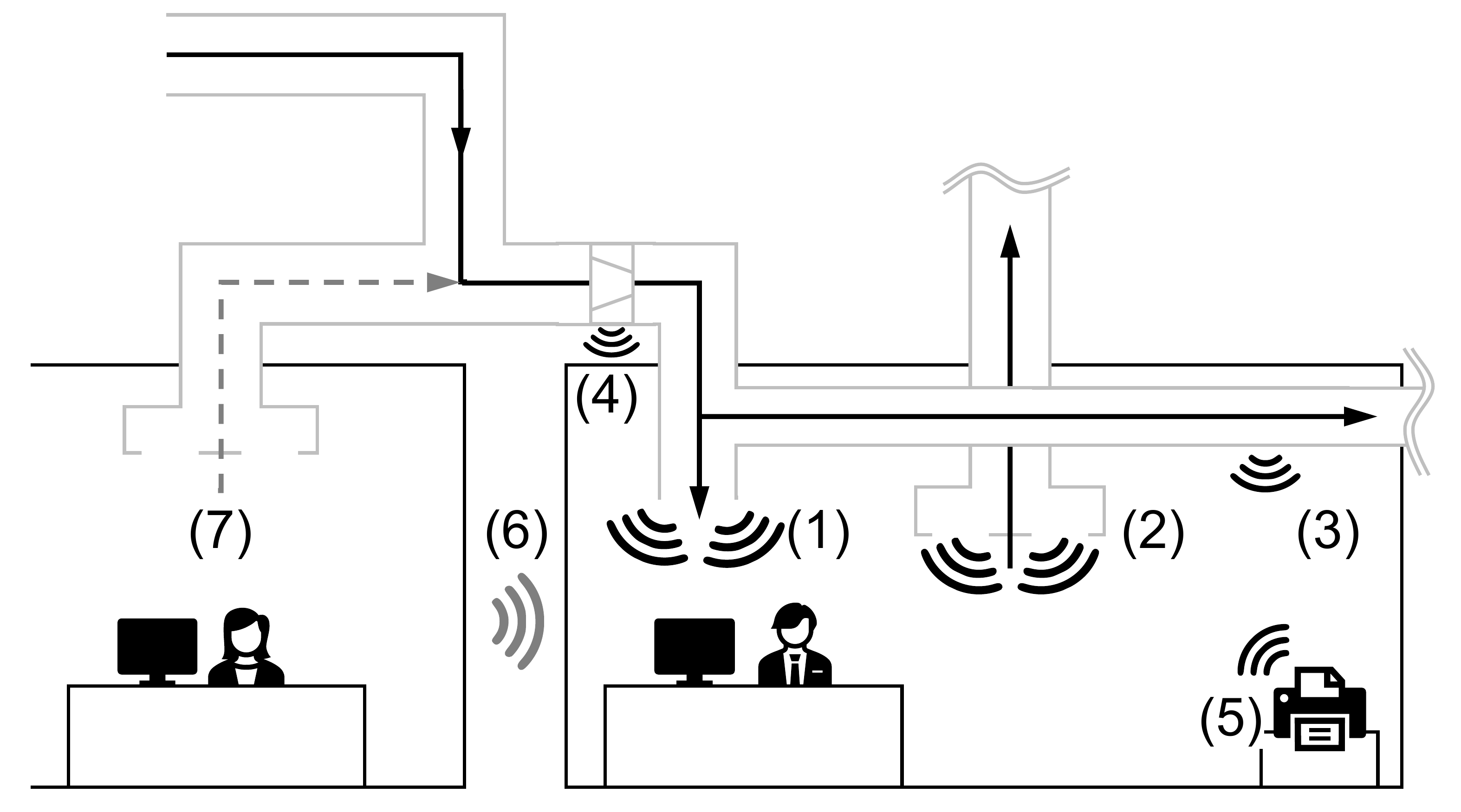}
	\caption{sources of noise in a room. (1) supply air system, (2) exhaust air system, (3) radiation noise of in-room ducts, (4) radiation noise of VFC/fans, (5) noise sources within the rooms, (6) noise transmission between rooms through walls and finally (7) cross-talk transmission between rooms}
	\label{fig:noise_sources}
\end{figure*}

Noise generation in a building comes from different sources. The level addition of all individual sources must not exceed noise limits. In VDI 2081, the noise in the rooms results from seven sources that are visualised in \cref{fig:noise_sources}:
\begin{enumerate}
    \label{it:supply} \item supply air system
    \label{it:exhaust} \item exhaust air system
    \item radiation noise of in-room ducts (or nearby)
    \item radiation noise of volume flow controllers / fans
    \label{it:inroom} \item noise sources within the rooms
    \label{it:cross_walls}\item noise transmission between rooms through walls
    \label{it:cross_talk} \item cross-talk sound transmission between rooms
\end{enumerate}

For every source of noise, the octave sound power level is converted into a octave sound pressure level by a conversion factor $l^{\mathrm{W}\rightarrow\mathrm{p}}$, A-weighted and level added. Finally, the A-weighted sound pressure level has to be lower than the noise limit $l^\mathrm{p,A, thresh}$:
\begin{align}
    \label{eq:spl_conversion}
    l^\mathrm{p,A} = 10 \logt{\sum_{f\in\mathcal{F}} 10^{0.1( \lw_f + l^{\mathrm{W}\rightarrow\mathrm{p}} + l^\mathrm{A})}} \leq l^\mathrm{p,A, thresh}.
\end{align}
For different types of noise, only the conversion factor from octave sound power level to octave sound pressure level $l^{\mathrm{W}\rightarrow \mathrm{p}}$ changes. In the following, the two conversions used in this manuscript are defined according to VDI 2081.

\paragraph{Airborne sound propagation into rooms}
\label{sec:airborne_noise}

The airborne sound power to sound pressure level conversion can be described according to VDI 2081 by the following model for rooms:
\begin{equation}
    \label{eq:airborne_sound_propagation}
    l^{\mathrm{W}\rightarrow \mathrm{p}} =  10\logt{\frac{Q {r^\mathrm{ref}}^2}{4 \pi {r^\mathrm{min}}^2} + \frac{4 n A^\mathrm{ref}}{A}}.
\end{equation}
The first factor is a location-related sound that includes a directional factor $Q$ and the distance from the outlet to the head of the nearest person $r^\mathrm{min}$. The next factor includes the equivalent absorption area $A$ and the number of equal outlets $n$. $r^\mathrm{ref}=\SI{1}{\meter}$ and $A^\mathrm{ref}=\SI{1}{\meter^2}$ are reference values for making the logarithm dimensionless. 

\paragraph{Sound radiation of in-room ducts}
\label{sec:inroom_noise}

The second source of noise in the rooms is the sound radiation of ducts in the room or nearby, whereby the sound is propagating through the duct walls.
In order to determine the octave sound pressure levels based on the octave sound power levels, VDI 2081 distinguishes between two factors: sound dampening due to the duct insulation and sound transmission into the room:
\begin{equation}
    \label{eq:sound_radiation}
    l^{\mathrm{W}\rightarrow \mathrm{p}} = R^\mathrm{ia} + 10\logt{\frac{S_k S^\mathrm{ref}}{S_1 A_2}} + K + 6,
\end{equation}
where $R^\mathrm{ia}$ is the dampening of an insulation to be determined. $S_k$ is the transmission area of the wall, $S_1$ the duct cross-sectional area, $A_2$ the equivalent sound absorption area of the room and $K$ the solid angle index as a measure for where the duct is located in the room. $S^\mathrm{ref}=\SI{1}{\meter^2}$ is a reference area for making the logarithm dimensionless.
\\
\\
\\
In this manuscript, not all types of noise are taken into account: The noise transmission between rooms through walls is not considered as it is part of the building's structural design. Furthermore, the cross-talk sound transmission between rooms is neglected as it is assumed that silencers in front of the rooms are sufficient. The radiation noise produced by the VFC or distributed fans is not explicitly considered in this manuscript. Instead, only sound-insulated components with low structure-borne noise design are used near the rooms. The sound-insulation is added to the costs of a component if necessary. Noise sources within the rooms are known a-priori and can simply be level added to the other noise levels. Lastly, supply and exhaust air system are not considered in the same model. The two systems can be solved individually which reduces the model size tremendously. Then, noise limits for both systems in sum should not exceed the overall noise limits per room.

The above mentioned approaches for ventilation system optimisation and acoustic modeling are combined in the following. While the airflow-only model contains rooms only as target pressure nodes, the coupled model needs to distinguish two more elaborate room models. In both cases, the resulting A-weighted sound pressure level has to fulfill a certain threshold, see \cref{eq:spl_conversion}. While the conversion and the A-weighting terms can be calculated a-priori from the geometry, material and flow regimes the level addition has to be modelled. Therefore, the A-weighted octave sound pressure levels are level added. Thus, the linearisation from \cref{sec:lvl_add} is applied seven times for each room:
\begin{align}
\label{eq:seven_level_additions}
    l^\mathrm{p,A} = \left(\left[\left\{l^\mathrm{p,A}_{f=1} ~ \tilde{+} ~ l^\mathrm{p,A}_{f=2} \right\} ~\tilde{+}~ l^\mathrm{p,A}_{f=3} \right] ~~ \dots ~~ \tilde{+} ~l^\mathrm{p,A}_{f=8} \right).
\end{align}

\subsubsection{Level addition}
\label{sec:lvl_add}

\begin{figure}[htbp]
\centering
\includegraphics[width=0.5\textwidth]{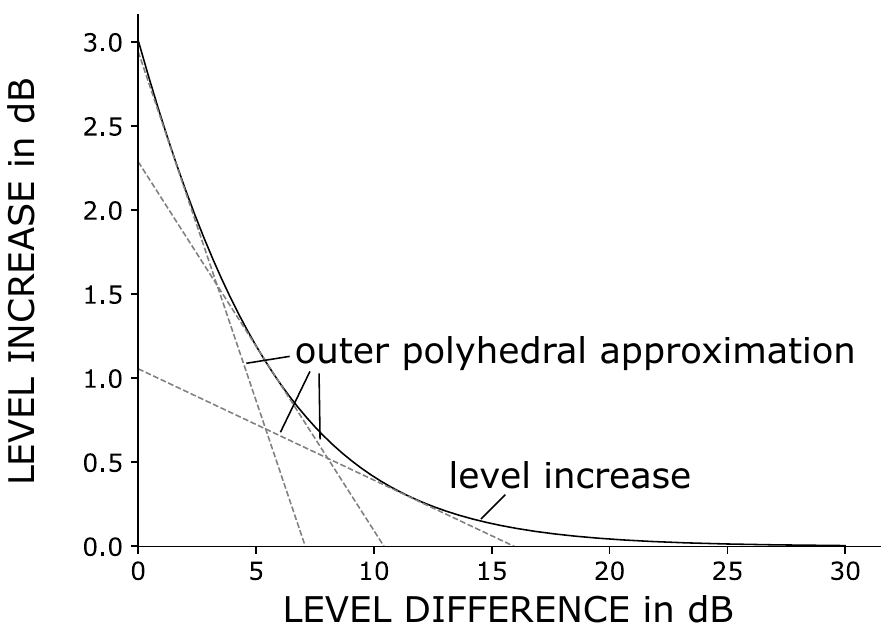}
\caption{level increase due to level addition and its outer polyhedral approximation. The level increase depends solely on the difference between the two levels. Figure taken from Breuer et~al.\cite{Breuer2024}} \label{fig:simp_lvl_add}
\end{figure}

Level addition, \cref{eq:level_add}, occurs eight times - due to eight octave sound power levels - per load case in every component for adding the dampened input octave sound power levels and the flow noises. Yet, it is a highly non-linear operation. As shown in Breuer et~al.~\cite{Breuer2024} it can be linearised using the fact that the level increase between any two levels depends solely on the absolute difference between them. The level increase as a function of level difference is still non-linear but can be approximated using polyhedral approximation, as shown in \cref{fig:simp_lvl_add}. The maximal error of the shown approximation is \SI{0.11}{\dB}.

The result is a linearised formulation of \cref{eq:level_add}, which for an element $(i,j)\in\mathcal{E}$ can be written as
\begin{align}
    \lwx{\mathrm{max}}_{f,i,j} = \max \left(\lw_{f,i} - d^\mathrm{W}_{f,i,j} , ~ \lwx{\mathrm{flow}}_{f,i,j} \right) &\qquad \forall f\in\mathcal{F} \quad \forall (i,j) \in \mathcal{C}\\
    \Delta \lw_{f,i,j} = 2 \lwx{\mathrm{max}}_{f,i,j} - (\lw_{f,i} - d^\mathrm{W}_{f,i,j}) - \lwx{\mathrm{flow}}_{f,i} &\qquad \forall f\in\mathcal{F} \quad \forall (i,j) \in \mathcal{C}\\
    \label{eq:lvl_add_too_simple} \lw_{f,j} = \lwx{\mathrm{max}}_{f,i,j} + \lwx{\mathrm{inc}}_{f,i,j} &\qquad \forall f\in\mathcal{F} \quad \forall (i,j) \in \mathcal{C}\\
    \lwx{\mathrm{inc}}_{f,i,j} \geq t_n(\Delta \lw_{f,i,j}) &\qquad \forall f\in\mathcal{F} \quad \forall (i,j) \in \mathcal{C} \quad \forall n \in {1,2,3}\\
    t_n = m_n \Delta \lw_{f,i,j} + b_n &\qquad \forall f\in\mathcal{F} \quad \forall (i,j) \in \mathcal{C} \quad \forall n \in {1,2,3}
\end{align}
with the maximum sound power level $\lwx{\mathrm{max}}_{f,i,j}$, the absolute difference between the two sound power levels $\Delta \lw_{f,i,j}$ and the linear equations $t_n$ with according to the outer polyhedral approximation.

The maximum is modelled according to Suhl~\cite{suhl2009}, introducing binary variables $z_{i,j,f}$:
\begin{align}
\lwx{\mathrm{max}}_{f,i,j} &\geq \lw_{f,i} - d^\mathrm{W}_{f,i,j} &\qquad \forall f\in\mathcal{F} \quad \forall (i,j) \in \mathcal{C}\\
    \lwx{\mathrm{max}}_{f,i,j} &\geq \lwx{\mathrm{flow}}_{f,i,j} &\qquad \forall f\in\mathcal{F} \quad \forall (i,j) \in \mathcal{C}\\
    \lwx{\mathrm{max}}_{f,i,j} - \lw_{f,i} - d^\mathrm{W}_{f,i,j} &\leq \lwx{\mathrm{max}} z_{i,j,f} &\qquad \forall f\in\mathcal{F} \quad \forall (i,j) \in \mathcal{C}\\
    \lwx{\mathrm{max}}_{f,i,j} - \lwx{\mathrm{flow}}_{f,i,j} &\leq \lwx{\mathrm{max}} (1-z_{i,j,f}) &\qquad \forall f\in\mathcal{F} \quad \forall (i,j) \in \mathcal{C}.
\end{align}

\cref{eq:lvl_add_too_simple} neglects that a change in octave sound power level can only be applied if the component is active (in case of fan, fan station or VFC) or purchased (in case of a silencer). Thus, \cref{eq:lvl_add_too_simple} is reformulated as follows:\begin{align}
    \lw_{f,j} - ( \lwx{\mathrm{max}}_{f,i,j} + \lwx{\mathrm{inc}}_{f,i,j}) & ~\substack{\leq + \\ \geq -} ~\lwx{\mathrm{max}}
    \begin{cases}
        (1-x_{i,j}), \text{if fan, fan station or VFC}\\
        (1-y_{i,j}), \text{if silencer}
    \end{cases}
    \qquad &\forall f \in \mathcal{F}\\
    \lw_{f,j} - \lw_{f,i} & ~\substack{\leq + \\ \geq -} ~ \lwx{\mathrm{max}} \begin{cases}
        x_{i,j}, \text{if fan, fan station or VFC}\\
        y_{i,j}, \text{if silencer}
    \end{cases}
    \qquad &\forall f \in \mathcal{F}.
\end{align}
    
\subsection{Solving algorithm}
\label{sec:solving_algorithm}

\begin{figure}[htbp]
\centering
\includegraphics[width=0.7\textwidth]{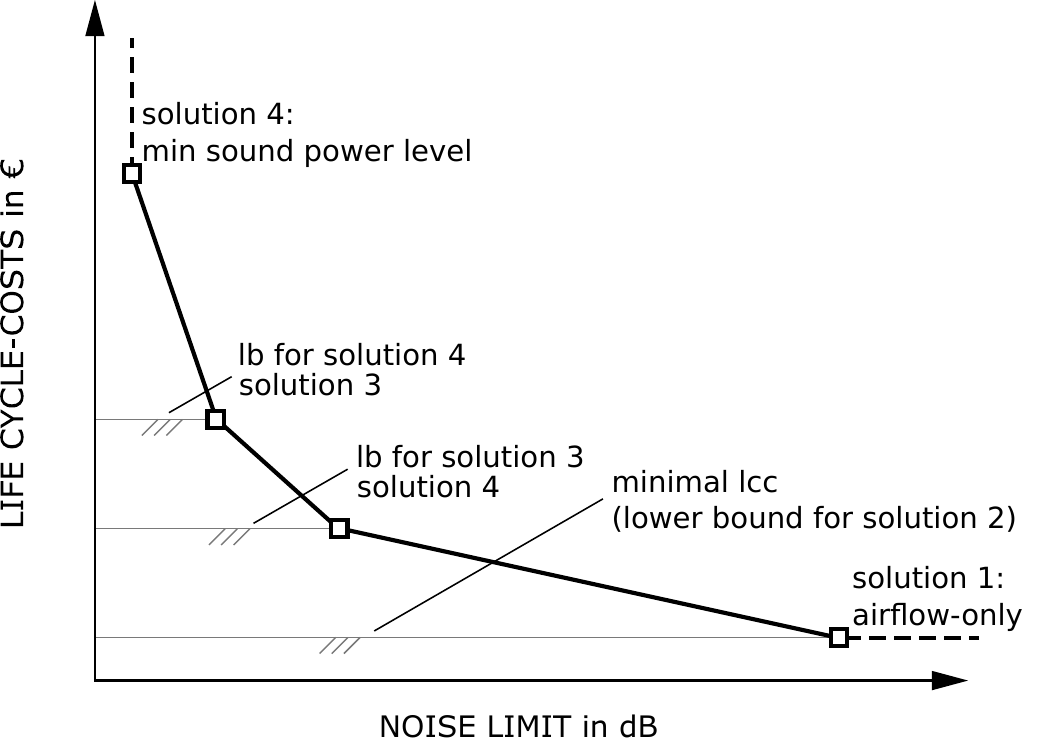}
\caption{visualisation of solving algorithm. The solving algorithm is used in an example to obtain a Pareto-front for the noise limits and the life-cycle costs. The solution 1, only regarding airflow, can be used as a lower bound (lb) to obtain a solution 2 with noise limits of $\SI{50}{\decibel}$. Then, solution 2 can be used as lb for obtaining solution 3, etc. This algorithm can be repeated until the noise limits allow no feasible solution. The last obtained solution provides the minimum feasible noise limit} \label{fig:solution_process}
\end{figure}

Compared to the airflow-only problem, the coupled optimisation problem contains far more variables and constraints, as will be shown in the case study - see \cref{sec:results}. This increases the computational effort to solve until global optimality tremendously. However, the problem's structure, i.e. the one-sided coupledness, can be exploited: The problem that neglects acoustics can be seen as an edge case of a system with infinitely high noise limits. This system can be optimised and its solution life-cycle costs can be used as lower limits -- a coupled system can never be cheaper than one that neglects acoustics. This lower bound helps the optimisation solver, especially the branch-and-bound algorithm, to find much better lower bounds in less time. This procedure can be iterated as the previous solution can always be set as a lower bound, as shown in \cref{fig:solution_process}. This algorithm allows for a more efficient calculation of the Pareto-front. The resulting Pareto-front allow decision makers a transparent decision making. In this manuscript, the decision between lower noise limits and lower life-cycle costs is investigated.

%% file: component_models.tex
\section{Component models}
\label{sec:component_models}
The ventilation system model contains models of components. In the following, five characteristic equations for \emph{fans}, \emph{volume flow controllers} (VFCs) and \emph{silencers} are given for the respective variables pressure change, power consumption, costs, dampening and flow noise:
\begin{equation}
    \text{component characteristic equations} : \{ \Delta p, \, \Pel, \,c, \,d_f, \,\lwx{\mathrm{flow}} \}.
\end{equation}
While VFCs and silencers always appear individually, multiple fans can be operated in parallel in a \emph{fan station}. Modelling the fans within a fan station - as a nested component - reduces the modelling effort and makes the model more flexible. Furthermore, a model for \emph{fixed components}, e.g. duct bendings or fire dampers, is given, which become necessary for the acoustic optimisation.

In engineering problems, model equations commonly exhibit non-linearity. To ensure the optimisation problem is solvable within reasonable timeframes, it is crucial to minimise non-linearity whilst maintaining accurate representation of underlying component data. As a result, novel model equations have been derived specifically for the VFC and silencer.

Parameters and constraints are defined for the four components. The sets are given in \cref{tab:set}, the component parameters in \cref{tab:comp_param} and the variables in \cref{tab:comp_vars}. 
Component parameters and variables for the acoustics are listed separately in \cref{tab:comp_param_acoustics}.

For modeling the component activation and purchasement decision, indicator variables are needed. All components have a binary variable indicating whether the component is bought. Fan stations, fans and VFCs also have an activation indicator that indicates whether the component is active in a certain scenario. These two indicator types are connected, as only purchased components can be active \cref{eq:activity_and_purchasement}.

\begin{align}
    \label{eq:activity_and_purchasement}
    x_{s,i,j} &\leq y_{s,i,j} \qquad &\forall s \in\mathcal{S} \quad \forall (i,j) \in \mathcal{C}
\end{align}

In the component models, Greek letters are used as regression coefficients in the model equations with a component-specific superscript added. Indices are written as subscripts, labels (e.g. '$\mathrm{ref}$' as in reference) as superscripts. For better readability, the five model equations and additional constraints for every component are introduced without indices $(i,j)\in\mathcal{C}$ for the respective edge - for the fans the indices $(p,d,n)\in\mathcal{C}^\mathrm{fan}$ are also dropped.

\subsection{Fans}
\label{sec:fan_model}
The key components for good air quality in the rooms are fans. Although essential for delivering volume flow and increasing pressure, they also have high power consumption and are the primary source of noise. Additionally, fans do not offer any form of dampening.
To account for the variety of fans in the market, two fan product lines $\mathcal{P}$ are used to create two individual fan models. Within a fan product line, fans are similar, meaning that the blade dimension ratios, the velocity triangles and the force ratios in the flow are the same~\cite{Carolus.2020}. Hence, the fans characteristic equations can be scaled depending on the fan diameter $D$. The different product lines are indicated by the subscript $p\in \mathcal{P}$. 

All fans are modelled with polynomial approximations for the pressure-volume flow characteristics, the power-volume flow characteristics and the cost characteristic. For the octave sound power level characteristics, logarithms are used.

The approach for approximating the pressure and the power consumption is based on scaling laws~\cite{eck1972ventilatoren} as previously demonstrated e.g. in Müller et al.~\cite{mueller2022}. The pressure increase $\Delta p = f(q,n)$ is modelled using a quadratic approximation which relates the pressure increase $\Delta p$ to the variable volume flow $q\in [\underline{q},\overline{q}]\subseteq \mathbb{R_0^+}$, the variable rotational speed $n\in [\underline{n},\overline{n}]\subseteq \mathbb{R_0^+}$ and the diameter $D$, see \cref{eq:fan_dp}. By normalising the rotational speed using its maximum value, the upper bound is $\overline{n}=1$.
The electric power consumption $\Pel = g(q,n,d)$ is modelled by a cubic approximation, see \cref{eq:fan_pel}.

The investment costs for each fan product line are modelled linearly with its diameter, \cref{eq:fan_c}. Additionally, if the fans are directly infront of rooms, $X^\mathrm{clad}=1$, sound-insulation costs are added, see \cref{sec:method_overview}. This is valid for fans in the same product line.

\begin{figure}[htpb]
     \centering
     \begin{subfigure}[b]{0.45\textwidth}
        \centering
    	\includegraphics[width=\textwidth]{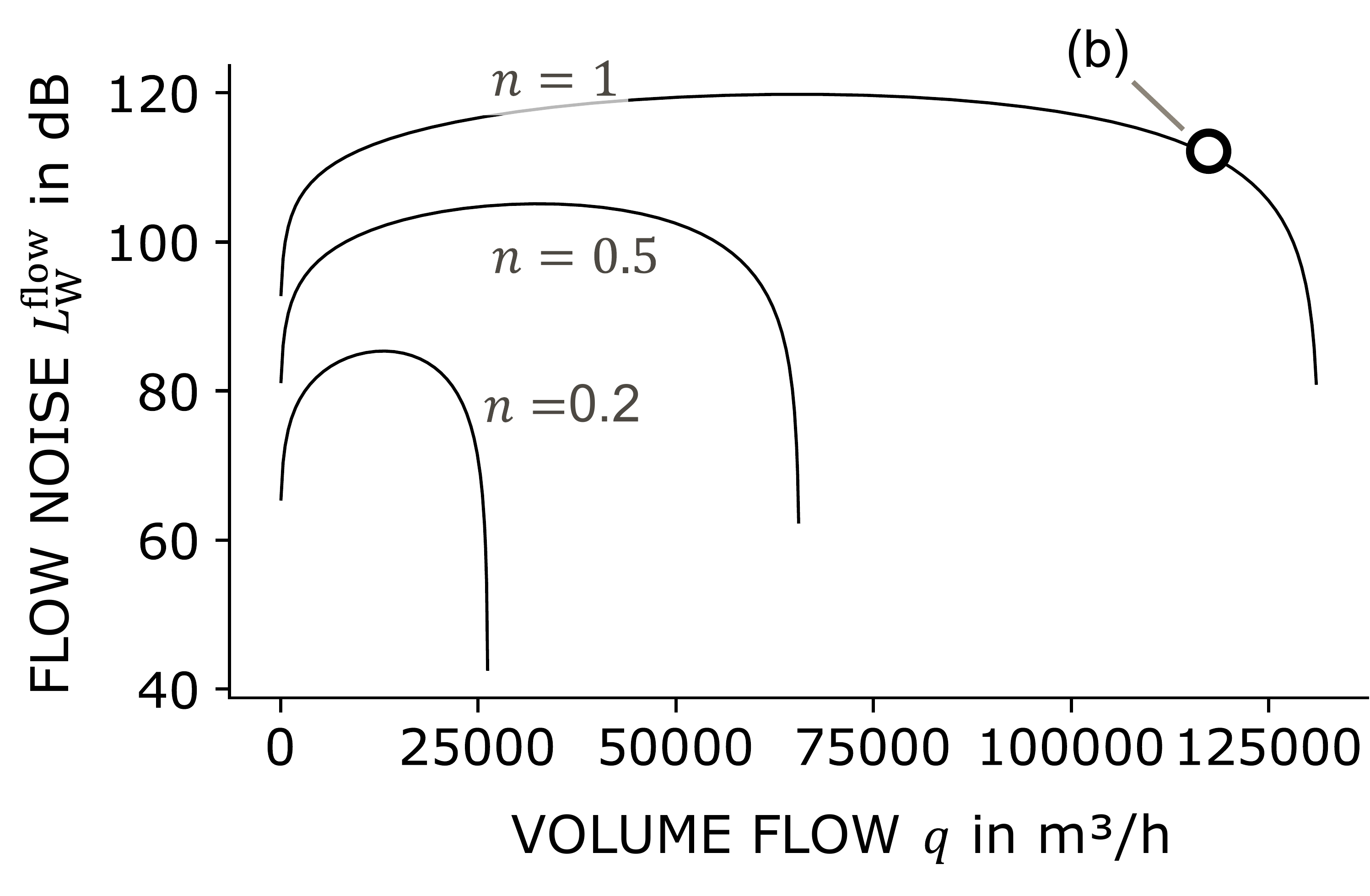}
    	\caption{flow noise characteristic $\lwx{\mathrm{flow}}$ of an example fan for three different rotational speeds.  The flow noise curve consists of level added octave flow noises shown in (b).}
    	\label{fig:flow_noise_char_tot}
     \end{subfigure}
     \hfill
     \begin{subfigure}[b]{0.45\textwidth}
         \centering
         \includegraphics[width=\textwidth]{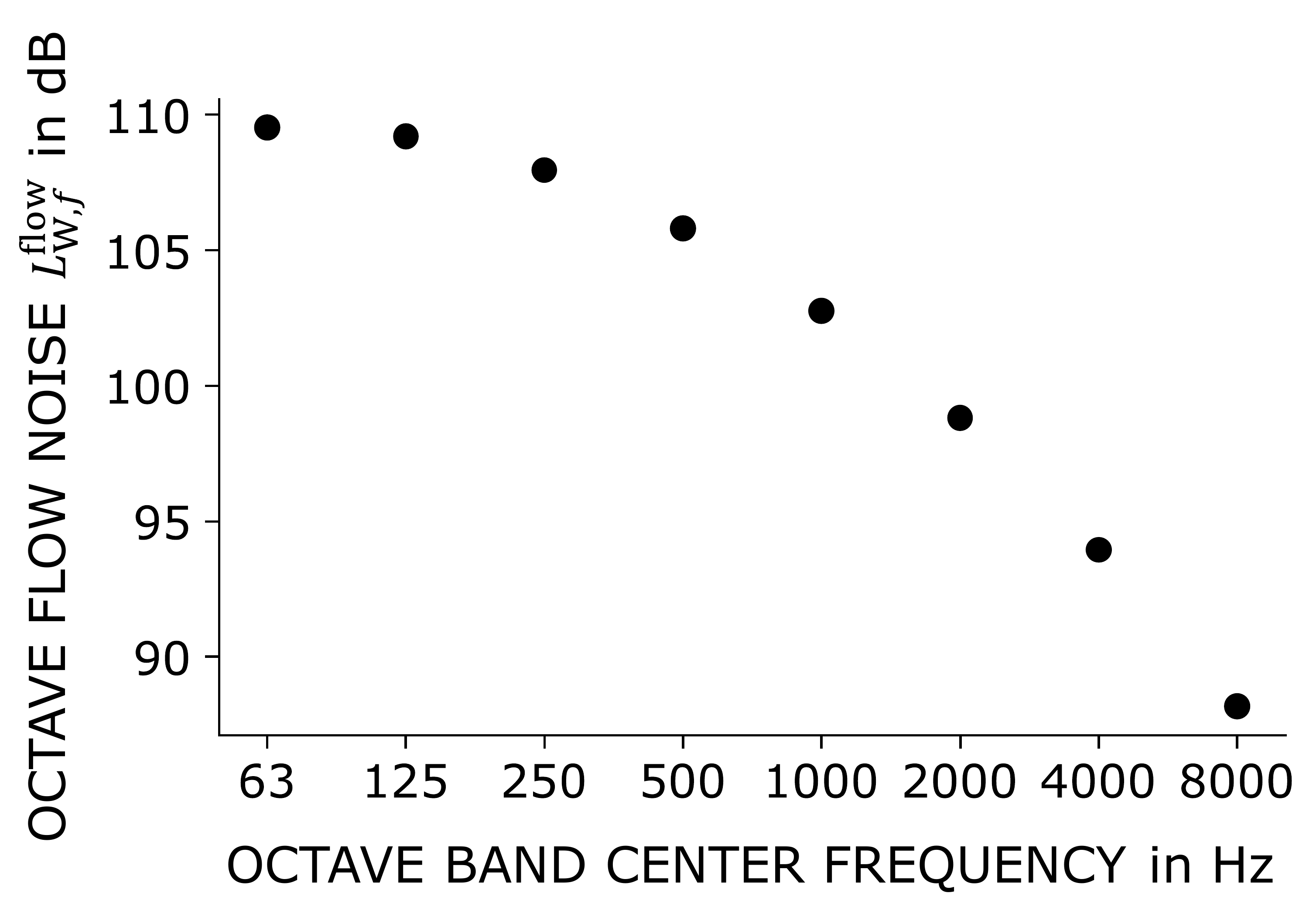}
         \caption{octave flow noise characteristic $\lwx{\mathrm{flow}}_f$ of a fan.}
         \label{fig:flow_noise_char_f}
     \end{subfigure}
     \caption{flow noise characteristics for an example fan.}
    \label{fig:flow_noise}
\end{figure}

The fan induces not dampening, \cref{eq:fan_d}. For obtaining the fan's flow noise $\lwx{\mathrm{flow}}_f = h(q,n,\Delta p,f)$, the fan is modelled according to Madison~\cite{madison1949fan}, as proposed in VDI 2081, see \cref{eq:fan_lw}. An example for the flow noise characteristic $\lwx{\mathrm{flow}}_f$ is shown in \cref{fig:flow_noise}. The term $\lwx{\mathrm{S}}(f,D,n)$ is a non-linear function of the rotational speed $n$, the diameter $D$ and the octave band $f\in\mathcal{F}$. To simplify the fan characteristics, this term is approximated by a polynomial, while the dependence on the volume flow and pressure is still represented by a logarithm, see \cref{eq:fan_lws}. This has proven to be necessary in order to ensure the most accurate approximation possible with low non-linearity.

Each model equation's coefficients of determination also called $R^2$-scores are calculated. If multiple approximations exist for the model equation, e.g. two fan product lines or octave sound power levels, the mean value is given. The resulting $R^2$-scores for the fans and all other component model equations are shown in \cref{tab:r2_scores}.

The five model equations that describe the fan characteristics are
\begin{align}
    \label{eq:fan_dp}\Delta p &= \sum_{m=0}^2 \alpha^\mathrm{fan}_{p,m} q^{2-m} n^m D^{-4+3m} \\
    \label{eq:fan_pel}\Pel &= \sum_{m=0}^3 \beta^\mathrm{fan}_{p,m} q^{3-m} n^m D^{-4+3m} + \beta_{p,4}^\mathrm{fan}\\
    \label{eq:fan_c} c &= \gamma_{p,0}^\mathrm{fan} D + \gamma_{p,1}^\mathrm{fan} + X^\mathrm{clad} \left(\gamma_{p,2}^\mathrm{fan} D + \gamma_{p,3}^\mathrm{fan}\right) \\
    \label{eq:fan_d} d_f & = 0
    \\
    \label{eq:fan_lw}\lwx{\mathrm{flow}}_f &= \lwx{\mathrm{S}} + 10 \logt{\frac{q}{q_\mathrm{ref}}} + 20 \logt{\frac{\Delta p}{p_\mathrm{ref}}} + \Delta \lwx{\mathrm{Oct}} - \Delta \lwx{\mathrm{Oct,tot}}  \qquad \forall f\in\mathcal{F}\\
    \label{eq:fan_lws} &= 10 \logt{\frac{q}{Q_\mathrm{ref}}} + 20 \logt{\frac{\Delta p}{P_\mathrm{ref}}} + \varepsilon^\mathrm{fan}_{p,0} n + \varepsilon^\mathrm{fan}_{p,1} n^2 + \varepsilon^\mathrm{fan}_{p,2} D + \varepsilon^\mathrm{fan}_{p,3} \qquad \forall f\in\mathcal{F},
\end{align}
with the reference parameters $Q^\mathrm{ref}=\SI{1}{\meter^3\per\second}$ and $P^\mathrm{ref} = \SI{1}{\pascal}$ used to make the logarithm dimensionless.

The coefficients $\alpha^\mathrm{fan}_{p,m}$, $\beta^\mathrm{fan}_{p,m}$ and $\gamma^\mathrm{fan}_{p,m}$ are obtained from regressing manufacturer provided fan data. Coefficients $\varepsilon^\mathrm{fan}_{p,m}$ are obtained by fitting the formulation of Madison for the different product lines.

\begin{table}[htbp]
\centering
\caption{Coefficients of determination $R^2$-score for component model equations}
\label{tab:r2_scores}
\begin{tabular}{  l l r }
  component & model equation  & $R^2$-score \\ 
  \hline
  fan & $\Delta p$ & 0.993 \\
  & $\Pel$ & 0.999\\
  & $c$ & 0.920 \\
  & $\lwx{\mathrm{flow}}_f$ & 0.946 \\
  vfc & $c$ & 0.975\\
  & $\lwx{\mathrm{flow}}_f$ & 0.966 \\
  silencer & $\Delta p $& 0.999\\
  & $c$ & 0.933 \\
  & $d^\mathrm{W}_{f}$ & 0.968 \\
  & $\lwx{\mathrm{flow}}_f$ & 0.998 \\
\end{tabular}
\end{table}

Unlike the other components, fans are always nested inside a fan station. The constraints connecting the fans to their fan station are given in \cref{sec:fs_model}. Fans differ in their product line $p\in\mathcal{P}$ and their diameter $d\in\mathcal{D}$.

To incorporate fans, the model equation for the power consumption, \cref{eq:fan_pel}, has to be reformulated by adding an intermediate variable, \cref{eq:po_model1}. This assures that the power consumption is only added to the electric energy consumption of the system if the fan is active, \cref{eq:po_model2}. Another constraint simplifies the topology decision making from the solver by ensuring an inactive fan does not contribute to the system's power consumption, \cref{eq:po_model3}.
\begin{align}
    \label{eq:po_model1} po^\mathrm{model} = \sum_{m=0}^3 \beta^\mathrm{fan}_{p,m} q^{3-m} n^m D^{-4+3m} + \beta_{p,4}^\mathrm{fan}\\
    \label{eq:po_model2} \Pel - po^\mathrm{model} \pml \overline{po}~(1-x)\\
    \label{eq:po_model3} \Pel \leq \overline{po}~x
\end{align}

\subsection{Fan Stations}
\label{sec:fs_model}

The fan station can accommodate either a single fan or multiple fans in parallel. They are placed on the respective edges $C^\mathrm{fs}$. A fan in the fan station is denoted by $(p,d,n) \in \mathcal{C}^\mathrm{fan} \subseteq (\mathcal{P}\times\mathcal{D}\times\mathcal{N})$ with product line $p \in \mathcal{P}$, diameter $D\in\mathcal{D}$ and $n \in \mathcal{N}$ for enumerating multiple identical fans. When arranged in parallel, fans provide an equal increase in pressure. The total volume flow through the fan station is determined by the load scenario, with the combined volume flows of the individual fans matching the station's total volume flow $Q$, as expressed in the equation:
\begin{equation}
    \label{eq:sumq}
    \sum_{(p,d,n) \in C^\mathrm{fan}} q_{p,d,n} = Q.
\end{equation}

The power consumption and investment costs for the fan station are calculated as the sum of those of their individual fans:
\begin{align}
    \Pel = \sum_{(p,d,n) \in C^\mathrm{fan}} \Pel_{p,d,n}\\
    c = \sum_{(p,d,n) \in C^\mathrm{fan}} c_{p,d,n}
\end{align}

To be able to connect fans and fan station in a linear formulation, an intermediate fan volume flow variable $q^\mathrm{fan}_{p,d,n}$ is defined in the fan station. To differ between the intermediate and the final volume flow, the final volume flow is denoted as $\mathrm{fan}_{p,d,n}.q$. This is done identically for other properties. The sum of the intermediate fan volume flows is equal to the fan station's volume flow -- which is a constant, \cref{eq:continuity1}. The intermediate fan volume flow, however, is only equal to the fan volume flow if the fan is active -- it is zero otherwise, \cref{eq:continuity2}. To simplify topology decisions for the solver another constraint is added: the intermediate volume flow is only non-zero if the fan is active, \cref{eq:continuity3}.

In addition to the constraints connecting the fan station pressure to the system, \cref{eq:energy_conservation1,eq:energy_conservation2,eq:energy_conservation3,eq:energy_conservation4}, the fan station pressure has to be connected to its fans. Thus, for each fan station, constraints connect every fan's pressure increase to the fan station's pressure increase if the fan is active, see \cref{eq:pressure_fan_to_fan_station1}. Again, to simplify topology decisions for the solver, the fan station's pressure increase is zero if the fan station is inactive, \cref{eq:pressure_fan_to_fan_station2}.

For the topology, fans can only be active if they are purchased \cref{eq:fan_purchasement}. Furthermore, a fan can only be purchased if the corresponding fan station is purchased, \cref{eq:fan_purchasement2}. The number of fans in a fan station is limited by $N^{\mathrm{max}}$,  \cref{eq:limit_N_max}.

\begin{align}
    \label{eq:continuity1}\sum_{(p,d,n)\in\mathcal{C}^\mathrm{fan}}q^\mathrm{fan}_{p,d,n} &= Q\\
    \label{eq:continuity2}q^\mathrm{fan}_{p,d,n} - \mathrm{fan}_{p,d,n}.q &\pml \mathrm{fan}_{p,d,n}.\overline{q} ~ (1-\mathrm{fan}_{p,d,n}.x) \qquad &\forall (p,d,n) \in \mathcal{C}^\mathrm{fan}\\
    \label{eq:continuity3}q^\mathrm{fan}_{p,d,n} &\leq \mathrm{fan}_{p,d,n}.\overline{q} ~ \mathrm{fan}_{p,d,n}.x \qquad &\forall (p,d,n) \in \mathcal{C}^\mathrm{fan}\\
    \label{eq:pressure_fan_to_fan_station1} (p^\mathrm{out} - p^\mathrm{in}) - \mathrm{fan}_{p,d,n}.\Delta p &\pml \overline{\Delta p}~(1-\mathrm{fan}_{p,d,n}.x) \qquad &\forall (p,d,n) \in \mathcal{C}^\mathrm{fan}\\ 
    \label{eq:pressure_fan_to_fan_station2} (p^\mathrm{out} - p^\mathrm{in}) &\pml \overline{\Delta p}~x \qquad &\forall (p,d,n) \in \mathcal{C}^\mathrm{fan}
    \\
    \label{eq:fan_purchasement} \mathrm{fan}_{p,d,n}.x &\leq \mathrm{fan}_{p,d,n}.y \qquad &\forall (p,d,n) \in \mathcal{C}^\mathrm{fan}
    \\
    \label{eq:fan_purchasement2} \mathrm{fan}_{p,d,n}.y &\leq y \qquad &\forall (p,d,n) \in \mathcal{C}^\mathrm{fan}\\
    \label{eq:limit_N_max}\sum_{(p,d,n) \in \mathcal{C}^\mathrm{fan}} y_{i,j,p,n} &\leq N^\mathrm{max}_{i,j} &\qquad \forall (i,j) \in \mathcal{C}^\mathrm{fs}.
\end{align}


For central systems that contain only one fan station $|\mathcal{C}^\mathrm{fs}|=1$, a lower bound for the electric power of the fan station can be obtained that has been shown to reduce the computation times tremendously, see Müller et~al.\cite{mueller2022}. Therefore, the electric power consumption of the central fan station is larger than product of the highest efficiency of an individual fan and the fan station's hydraulic power, see \cref{eq:phyd}. The hydraulic power has to be calculated for each scenario and weighted summed by the scenario's time share. In each scenario, the volume flow is known a-priori  and also for the pressure a sensible estimate can be done: The minimum needed pressure increase of the fan station can be calculated as the maximum of all room pressure requirements, neglecting pressure increases due to VFCs or silencers:
\begin{align}
    \label{eq:phyd} po_{\mathrm{hyd},s} = \overline{\eta}^\mathrm{fan} \max_{i \in \mathcal{V}_\mathrm{T}} \left(p_{i,s}^T \right)\sum_{(i,j)\in\mathcal{E}, j \in \mathcal{V}_{\mathrm{R}}} q_{s,i,j}  \leq \sum_{(i,j)\in\mathcal{C}^\mathrm{fs}}\sum_{(p,d,n)\in\mathcal{C}^\mathrm{fan}} \Pel_{i,j,p,d,n,s} \qquad \forall s \in \mathcal{S}.
\end{align}

Concerning acoustics, the fans input octave sound power levels are equal to the fan station's input levels, \cref{eq:fan_spl1}. On the other hand, the outlet fan station octave sound power levels of the fans must be equal to the level added sum of the octave sound power levels of all fans. This would, however, lead to significant coupling in presence of multiple fans and would increase the computation time tremendously. Thus, to simplify the octave sound power level of the fan station is equal to the maximum of the fans' octave sound power levels \cref{eq:fan_spl2}.
Finally, if the fan station is inactive, the inlet and outlet octave sound power levels are zero, \cref{eq:fs_fan_spl_off}.

\begin{align}
    \label{eq:fan_spl1}\lwx{\mathrm{in}}_{f} &= \mathrm{fan}_{p,d,n}.\lwx{\mathrm{in}}_{f}  \qquad &\forall f \in \mathcal{F} \quad \forall (p,d,n)\in\mathcal{C}^\mathrm{fan}\\
    \label{eq:fan_spl2}\lwx{\mathrm{out}} &\leq \mathrm{fan}_{p,d,n}.\lwx{\mathrm{out}}_{f} \qquad &\forall f \in \mathcal{F} \quad \forall (p,d,n)\in\mathcal{C}^\mathrm{fan}\\
    \label{eq:fs_fan_spl_off} \lwx{\mathrm{out}}_{f} - \lwx{\mathrm{in}}_{f} &\pml \overline{L} x \qquad &\forall f \in \mathcal{F}
\end{align}

\subsection{Volume Flow Controllers}
\label{sec:VFC_model}

\begin{figure}[htpb]
     \centering
     \begin{subfigure}[B]{0.45\textwidth}
        \centering
    	\includegraphics[trim= 0 -130 0 0, clip, width=0.9\textwidth]{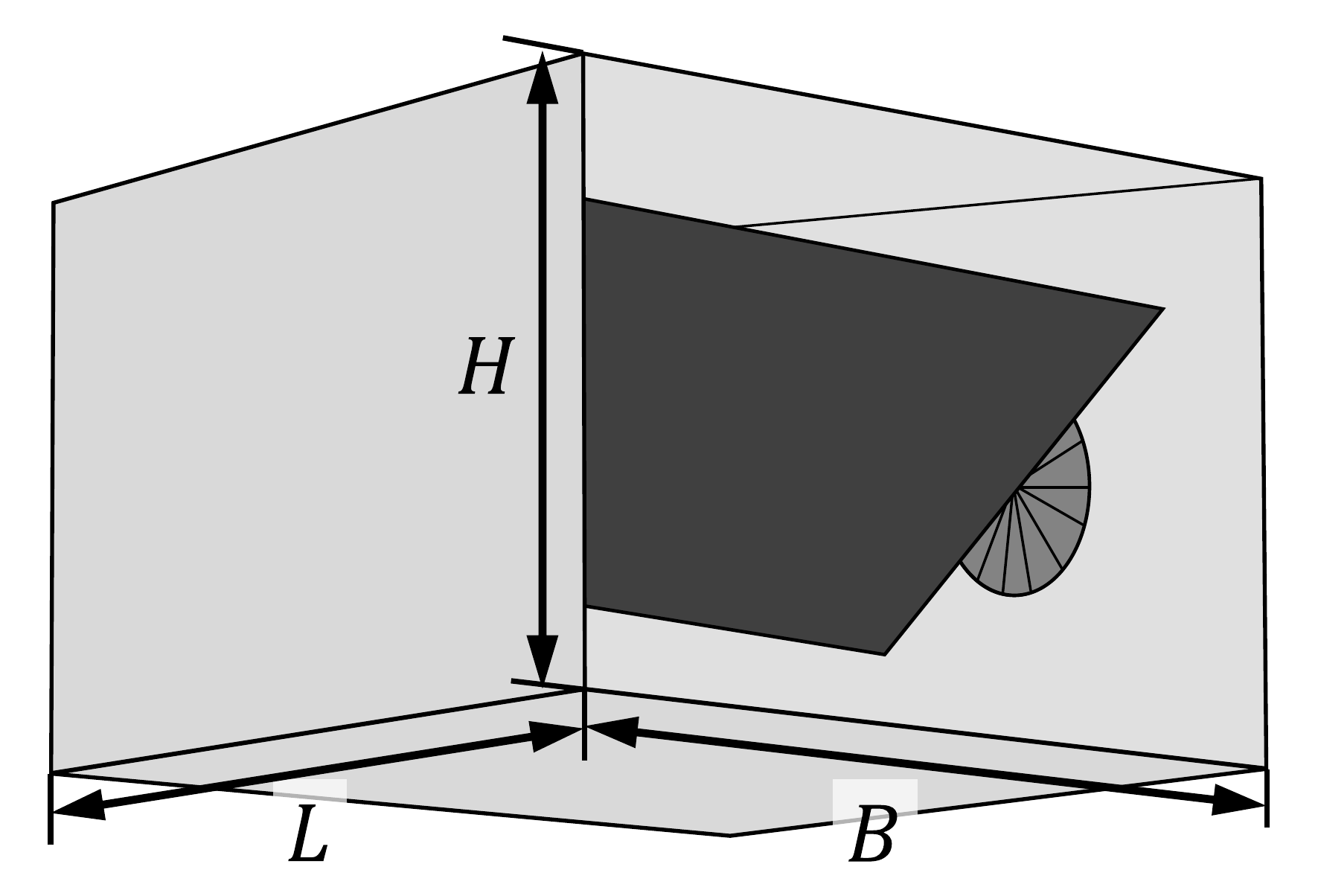}
    	\caption{rectangular volume flow controller}
    	\label{fig:VFC}
     \end{subfigure}
     \hfill
     \begin{subfigure}[b]{0.45\textwidth}
         \centering
         \includegraphics[width=0.9\textwidth]{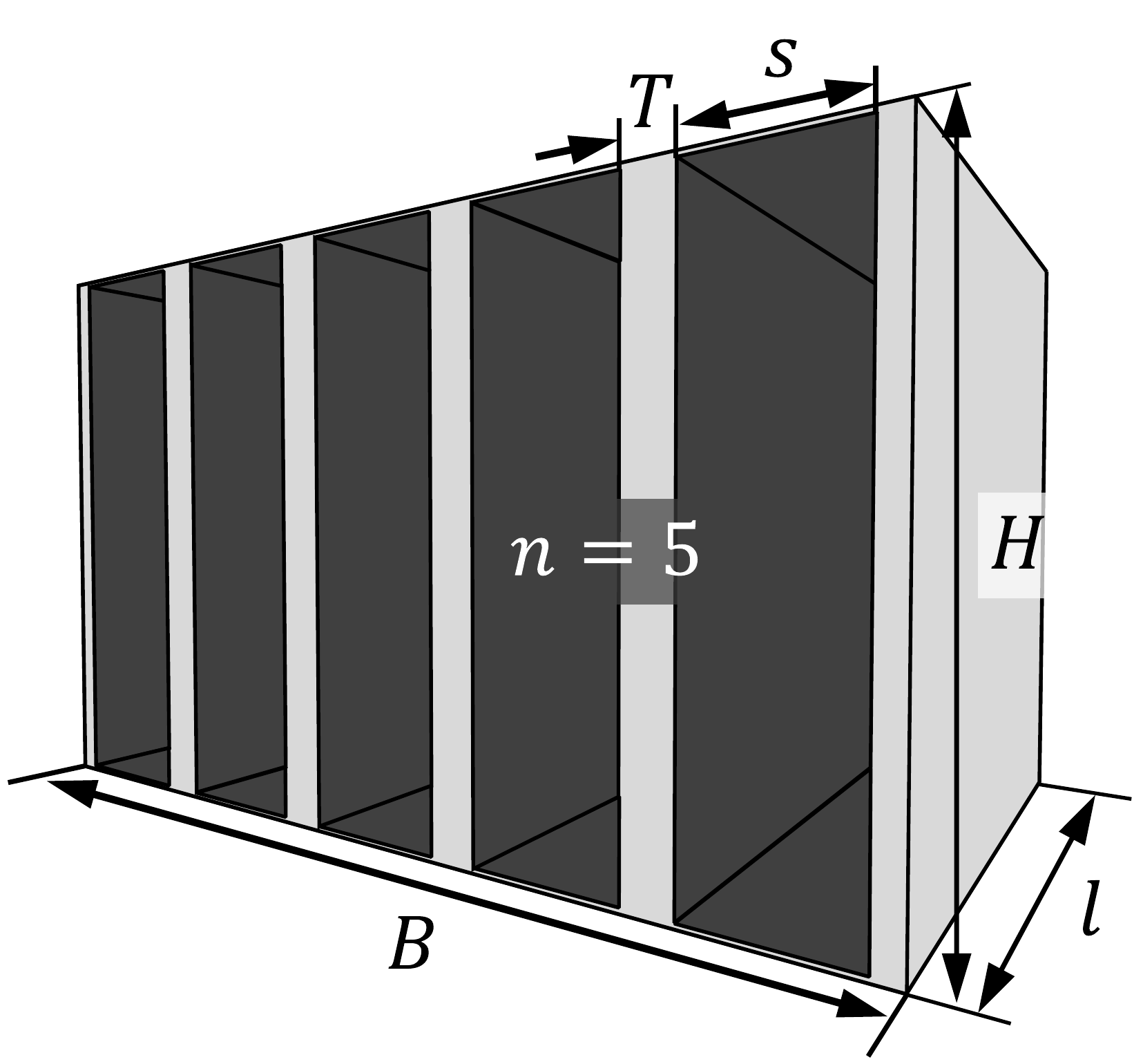}
         \caption{splitter silencer with $n=5$ splitter elements.}
         \label{fig:splitter_silencer}
     \end{subfigure}
     \caption{component representation with dimensioning}
    \label{fig:vis_VFC_sil}
\end{figure}

The volume flow controllers (VFC) are used to throttle the airflow by inducing a variable pressure loss in the duct. As the duct network of the ventilation system is predefined, the dimensioning - height $H$ and width $B$ - is also fixed, see \cref{fig:VFC}. In addition, the volume flow $Q$ is known for each VFC in each load scenario. Thus, the only variable in the VFC model is the pressure loss $\Delta p$. In this manuscript, only models for rectangular VFCs are presented, although similar formulas can be derived for circular VFCs.

In contrast to modelling multiple fan product lines, a single but general VFC product line is modelled. This type is generated based on data from multiple manufacturers. The length is not included, as all VFCs used for the modelling have the same length. The pressure loss can be set arbitrary within predefined bounds, $\Delta p \in [\underline{\Delta p},\overline{\Delta p}] \subseteq \mathbb{R}_0^+$. The electric energy consumption of the VFCs is negligible as energy is only used to change the throttling and not during the operation, \cref{eq:vfc_power}. The costs are modelled as a function of the VFC dimensions, they are constant for a given VFC, \cref{eq:vfc_cost}. They too are fitted from multiple manufacturers' data. Additionally, as pointed out in \cref{sec:acoustic_room_models} VFCs close to rooms will be sound-insulated which adds constant costs.

Furthermore, VFCs induce flow noise but offer no dampening. The flow noise $\lwx{\mathrm{flow}}_f=k(Q,H,B,\Delta p)$ is modelled by approximating manufacturer data, using linear approximation in the variable pressure, see \cref{eq:vfc_flownoise}. 

\begin{align}
    \label{eq:vfc_pressure} \Delta p &= \{\underline{\Delta p}, \overline{\Delta p}\}\\
    \label{eq:vfc_power} \Pel &= 0\\
    \label{eq:vfc_cost} c &= \gamma^\mathrm{vfc}_0 HB + \gamma^\mathrm{vfc}_1 + X^\mathrm{clad} \gamma^\mathrm{vfc}_2 = const.\\
    d_f &= 0\\
    \label{eq:vfc_flownoise} \lwx{\mathrm{flow}}_f &= \varepsilon^\mathrm{vfc}_{f,0} \Delta p + \varepsilon^\mathrm{vfc}_{f,1} \frac{Q^2}{(HB)^2} + \varepsilon^\mathrm{vfc}_{f,2} \frac{Q}{HB} + \varepsilon^\mathrm{vfc}_{f,3} (HB) + \varepsilon^\mathrm{vfc}_{f,4} \sqrt{HB} + \varepsilon^\mathrm{vfc}_{f,5} \qquad \forall f \in \mathcal{F}
\end{align}

The parameters $\varepsilon^\mathrm{vfc}_{f,m}$, $\gamma^\mathrm{vfc}_m$ and $\alpha^\mathrm{vfc}$ are derived from regressing manufacturer data.

For the VFC, the model equations from \cref{sec:VFC_model} are used. For the pressure loss, constraints have to assure the physical feasability: If the VFC is not active, the pressure loss is zero, \cref{eq:VFC_pressure2}. Additionally, the difference between inlet pressure and outlet pressure is equal to the pressure loss, \cref{eq:VFC_pressure3}.
\begin{align}
    \label{eq:VFC_pressure2} \Delta p &\leq \overline{\Delta p}~x\\
    \label{eq:VFC_pressure3} p^\mathrm{out} - p^\mathrm{in} &= - \Delta p 
\end{align}

\subsection{Silencers}

Silencers are used for reducing the octave sound power levels in the duct. Simultaneously, silencers introduce a pressure loss and flow noise. In the presented approach only splitter silencers are modelled and applied. However, similar models can be derived for duct silencers.

Silencers are dimensioned via the parameters height $H$, width $B$ and splitter width $T$ as well as the variables length $l\in [\underline{l},\overline{l}]\subseteq \mathbb{R}_0^+$ and number of splitter elements $n \in [\underline{n}, \overline{n}]\subseteq \mathbb{N}$, see \cref{fig:splitter_silencer}. For the purpose of modelling it helps defining two more variables, that are functions of the number of splitter elements.\\
The gap length $s$ can be purely derived from the geometry, see \cref{fig:splitter_silencer},
\begin{equation}
    B = n(T+s) \Leftrightarrow s = B/n-T.
\end{equation}
It can only assume discrete values.
Using the known volume flow $Q$, the gap velocity through the silencer can be calculated using the gap area
\begin{equation}
    v = \frac{Q}{A^\mathrm{gap}} = \frac{Q}{(sn)H} = \frac{Q}{(B-Tn)H}.
\end{equation}

The silencer models are placed on the respective edges, $(i,j) \in \mathcal{C}^\mathrm{sil}$. They are described using a set of variables and parameters based on manufacturer data. Therefore, manufacturer data is used. As model equations were hardly available in the literature and those available were highly non-linear, new model equations were derived based on manufacturer data. 

For the pressure loss a model equation from Rietschel et~al.\cite{Rietschel.1994} is used but simplified, see \cref{eq:sil_dp}. A linear cost model is obtained, again by approximating manufacturer data, see \cref{eq:sil_costs}. The silencer dampening $d^{\mathrm{W}}_f = d(n,l,B)$ is modelled using a quadratic approximation in $n$, see \cref{eq:sil_damp}.

As silencers also generate noise, the flow noise $\lwx{flow}=l(v,H,B)$ is modelled according to VDI 2081 Eq. 65. However, these contain logarithms of variables and thus a quadratic reformulation is used, see \cref{eq:sil_flow}.

\begin{align}
    \label{eq:sil_dp} \Delta p &= \alpha^\mathrm{sil}_0 v^2 + \alpha^\mathrm{sil}_1 \frac{v^2}{s} + \alpha^\mathrm{sil}_2 \frac{v^2 l}{s} \\
    \Pel &= 0\\
    \label{eq:sil_costs} c &= \gamma^\mathrm{sil}_0 n + \gamma^\mathrm{sil}_1 l B + \gamma^\mathrm{sil}_2 HB + \gamma_3 \\
    \label{eq:sil_damp} d^\mathrm{W}_{f} &= \delta^\mathrm{sil}_{f,0} n^2 + \delta^\mathrm{sil}_{f,1} B^2 n + \delta^\mathrm{sil}_{f,2} l n + \delta^\mathrm{sil}_{f,3} lB + \delta^\mathrm{sil}_{f,4} n + \delta^\mathrm{sil}_{f,5} l + \delta^\mathrm{sil}_{f,6} B + \delta^\mathrm{sil}_{f,7} \qquad \forall f \in \mathcal{F}\\
    \label{eq:sil_flow} \lwx{\mathrm{flow}}_f &= \varepsilon^\mathrm{sil}_{f,0} v^2 + \varepsilon^\mathrm{sil}_{f,1} v + \varepsilon^\mathrm{sil}_{f,2} HB + \varepsilon^\mathrm{sil}_{f,3} (HB)^2 + \varepsilon^\mathrm{sil}_{f,4} \qquad \forall f \in \mathcal{F}
\end{align}

The silencer's pressure loss only applies if the silencer is purchased, \cref{eq:sil_p_loss1}. Otherwise the pressure loss is zero, \cref{eq:sil_p_loss2}.
\begin{align}
    \label{eq:sil_p_loss1} p^\mathrm{in} - p^\mathrm{out} &\pml \overline{\Delta p} y\\
    \label{eq:sil_p_loss2} p^\mathrm{in} - p^\mathrm{out} - \Delta p &\pml \overline{\Delta p}~ (1-y)
\end{align}

While the costs of fans and VFCs are parameters, the silencer costs depend on variables length $l$ and number of splitter elements $n$. Thus, for the silencer costs an intermediate variable dependent on the geometry $c^\mathrm{geom}$ is defined. The final component costs $c$ are the intermediate silencer costs that apply only if the silencer is purchased, \cref{eq:sil_cost1}. Otherwise the costs are zero, \cref{eq:sil_cost2}.
\begin{align}
    c^\mathrm{geom} &= \gamma^\mathrm{sil}_0 n + \gamma^\mathrm{sil}_1 l B + \gamma^\mathrm{sil}_2 HB + \gamma_3 \\
    \label{eq:sil_cost1} c - c^\mathrm{geom} &\pml \overline{c}~(1-x)\\
    \label{eq:sil_cost2} c &\leq \overline{c}x
\end{align}

\subsection{Fixed components}
\label{sec:fixed_comp}

In addition to the above categories of components that are variables in the optimisation, the remaining components to be modelled are permanently installed and therefore part of the duct network. They do not contribute to costs or electric power consumption. Their pressure loss is calculated a-priori, - due to the linearity of pressure losses  - summed up and defined in the load cases. Hence, for fixed components, only flow noise and dampening need to be modelled. Flow noise and dampening for the fixed components can be obtained from VDI 2081 or from manufacturer data.

Breuer et~al. presented equations that combine multiple fixed elements occurring in series~\cite{Breuer2024}. This allows to reduce the number of elements in the duct network tremendously. Using this approach, the flow noise and attenuation of multiple fixed elements in series $\mathcal{B} \subset{E}$ - sorted ascending in flow direction - with source node $so$ and target node $ta$ are combined. The resulting octave flow noise $\hat{l}^\mathrm{W,flow,total}_{f,so,ta}$ and dampening $\hat{d}^\mathrm{W,total}_{f,so,ta}$ can be derived with the following formulas:
\begin{align}
    \hat{l}^\mathrm{W,flow}_{f,i,j} = l^\mathrm{W,flow}_{f,i,j} - \sum_{\substack{l=i+1,\\k = j+1}}^\mathcal{B} d^\mathrm{W}_{f,l,k} & \qquad   \forall (i,j) \in \mathcal{B} \quad \forall f \in \mathcal{F}\\
    \hat{l}^\mathrm{W,flow,total}_{f,so,ta} = \tilde{+}\sum_{(i,j)}^\mathcal{B} \hat{l}^\mathrm{W,flow}_{f,i,j} & \qquad \forall f \in \mathcal{F}\\
    \hat{d}^\mathrm{W,total}_{f,so,ta} = \sum_{(i,j)}^{\mathcal{B}} d^\mathrm{W}_{f,i,j} & \qquad \forall f \in \mathcal{F}.
\end{align}

\begin{table}[htbp]
\centering
\caption{Used parameters in component models. Respective edge indices $(i,j)$ (and $(p,d,n)$ for fans) are dropped for better readability.}
\label{tab:comp_param}
\begin{tabular}{  p{2cm} p{4cm} p{1cm} >{\raggedright}p{8cm} }
 component & parameter  & range & description \\ 
 \hline
 all & $\alpha_{m}$ & $\mathbb{R}$ & pressure regression coefficients for component $com\in \mathcal{C}$ with $m$ based on the number of regression coefficients. For fans, the subscript $(p,d)\in\mathcal{P}\times\mathcal{D}$ is added to account for the different product lines and diameters. \\
 & $\beta_{m}$ & $\mathbb{R}$ & power regression coefficients for component $com\in \mathcal{C}$ with $m$ based on the number of regression coefficients. For fans, the subscript $(p,d)\in\mathcal{P}\times\mathcal{D}$ is added to account for the different product lines and diameters.\\
 & $\gamma_{m}$ & $\mathbb{R}^+$ & cost regression coefficients for component $com\in \mathcal{C}$ with $m$ based on the number of regression coefficients. The subscript $p\in\mathcal{P}$ is only added for fans to account for the different product lines.\\
 & $\delta_{m}$ & $\mathbb{R}$ & sound dampening coefficients for component $com\in \mathcal{C}$ with $m$ based on the number of regression coefficients. For fans, the subscript $(p,d)\in\mathcal{P}\times\mathcal{D}$ is added to account for the different product lines and diameters.\\
 & $\varepsilon_{(p),m}$ & $\mathbb{R}^+$ & flow noise regression coefficients for component $com\in \mathcal{C}$ with $m$ based on the number of regression coefficients. For fans, the subscript $(p,d)\in\mathcal{P}\times\mathcal{D}$ is added to account for the different product lines and diameters.\\
   \hline
   fan & & & fan inside fan station $\mathcal{C}^\mathrm{fs}$
   \\ $(p,d,n)\in\mathcal{C}^\mathrm{fan}$ & $\underline{n}$, $\overline{n}$, $\underline{q}$, $\overline{q}$, $\underline{\Delta p}$, $\overline{\Delta p}$, $\underline{po}$, $\overline{po}$ & $\mathbb{R}_0^+$ & Upper $\overline{}$ and lower $\underline{}$ bounds for the operating parameter of the pump, i.e. rotational speed, volume flow, pressure increase and electric power consumption  \\
 & $D$ & $\mathbb{R}^+$ & fan diameter \\
 \hline
 fan station \\
 $(i,j)\in\mathcal{C}^\mathrm{fs}$ & $q$ & $\mathbb{R}^+$ & volume flow over the fan station edge \\
  & $N^\mathrm{max}$ & $\mathbb{N}$ & maximum number of fans in the fan station \\
  & $\overline{\Delta p}$ & $\mathbb{R}^+$  & Upper bound for the pressure at the fan station\\
  \hline
 splitter silencer \\
 $(i,j)\in\mathcal{C}^\mathrm{sil}$ & $q$ & $\mathbb{R}^+$ & volume flow over the silencer edge \\
 & $\underline{l}, \overline{l}, \underline{n}, \overline{n}, \underline{s}, \overline{s}, \underline{c}, \overline{c}, \underline{\Delta p}, \overline{\Delta p}$ & $\mathbb{R}_0^+$ & Upper $\overline{}$ and lower $\underline{}$ bounds for the silencer, i.e. length, number of splitter elements, gap width, silencer costs, pressure loss \\
 & $H$ & $\mathbb{R}^+$ & silencer height\\
 & $B$ & $\mathbb{R}^+$ & silencer width\\
 & $T$ & $\mathbb{R}^+$ & width of the gap between splitters\\
 \hline
 rect. VFC \\
 $(i,j)\in\mathcal{C}^\mathrm{vfc}$ & $q$ & $\mathbb{R}^+$ & volume flow over the silencer edge \\
 & $\overline{\Delta p}$ & $\mathbb{R}_0^+$ & Upper bound of the VFCs pressure loss\\
 & $H$ & $\mathbb{R}^+$ & VFC height\\
 & $B$ & $\mathbb{R}^+$ & VFC width\\
\end{tabular}
\end{table}

\begin{table}[htbp]
\centering
\caption{Used variables in component models. Respective edge indices $(i,j)$ are dropped for better readability.}
\label{tab:comp_vars}
\begin{tabular}{  p{2cm} p{2cm} p{1cm} >{\raggedright}p{9cm} }
 component & variable  & domain & description \\ 
 \hline
 all & & &\\
 & $p^\mathrm{in}_s$ & $[0,\overline{\Delta p}]$ & input pressure to the component in scenario $s\in\mathcal{S}$\\
 & $p^\mathrm{out}_s$ & $[0,\overline{\Delta p}]$ & output pressure of the component in scenario $s\in\mathcal{S}$\\
 & $x_s$ & $\{0,1\}$ & Binary variable indicating if component is active (not existent for silencers) in scenario $s\in\mathcal{S}$\\
 & $y$ & $\{0,1\}$ & Binary variable indicating if component is purchased \\
 \hline
 fan & &  & fan inside fan station $\mathcal{C}^\mathrm{fs}$ \\
 $(p,d,n)\in\mathcal{C}^\mathrm{fan}$ & $q_{p,d,n,s}$ & $[\underline{q},\overline{q}]$ & volume flow in scenario $s\in\mathcal{S}$\\
 & $n_{p,d,n,s}$ & [0,1] & rotational speed of fan $(p,d,n)\in\mathcal{C}^\mathrm{fan}$ in scenario $s\in\mathcal{S}$\\
 & $\Delta p_{p,d,n,s}$ & $[\underline{\Delta p}, \overline{\Delta p}]$ &  pressure increase of fan $(p,d,n)\in\mathcal{C}^\mathrm{fan}$ in scenario $s\in\mathcal{S}$\\
 & $\Pel_{p,d,n,s}$ & $[\underline{\Pel},\overline{\Pel}]$ & power consumption of fan $(p,d,n)\in\mathcal{C}^\mathrm{fan}$ in scenario $s\in\mathcal{S}$ \\
 \hline
 fan station & &  &  \\
  $(i,j)\in\mathcal{C}^\mathrm{fs}$ & $q^\mathrm{fan}_{p,d,n,s}$ & $[0,q]$ & volume flow of fan $(p,d,n)\in\mathcal{C}^\mathrm{fan}$ in scenario $s\in\mathcal{S}$ in fan station\\
 \hline
 splitter silencer & & & \\
 $i,j)\in\mathcal{C}^\mathrm{sil}$& $n$ & $[\underline{n},\overline{n}]$ & number of splitter elements\\
 & $l$ & $[\underline{l},\overline{l}]$ & length\\
 & $s$ & $[\underline{s},\overline{s}]$& gap width\\
 & $\Delta p_s $ & $[\underline{\Delta p}, \overline{\Delta p}]$ & pressure loss in scenario $s\in\mathcal{S}$ \\
 \hline
 rect. VFC & & & \\
 $(i,j)\in\mathcal{C}^\mathrm{vfc}$& $\Delta p_s$ & $[\underline{\Delta p}, \overline{\Delta p}]$ & pressure loss in scenario $s\in\mathcal{S}$ \\
 
\end{tabular}
\end{table}

\begin{table}[htbp]
\centering
\caption{Acoustic parameters and variables in all component models and fans. Component indices $(i,j)$ or for fans $(p,d,n)$ are dropped for better readability.}
\label{tab:comp_param_acoustics}
\begin{tabular}{ p{2cm} p{2cm} >{\raggedright}p{10cm} }
 parameter  & range & description \\ 
 \hline
 $\underline{\lw}$ & $\mathbb{R}^+$ & maximal sound power level of component\\
 $\overline{\lw}$ & $\mathbb{R}$ & minimal sound power level of component\\
 $m_i$ & $\mathbb{R}$ & slope of tangents used in outer polyhedral approximation, see \cref{sec:lvl_add}, $i\in\{1,2,3\}$\\
 $b_i$ & $\mathbb{R}$ & $y$-intercept of tangents used in outer polyhedral approximation, see \cref{sec:lvl_add}, $i\in\{1,2,3\}$\\
  $X^\mathrm{clad}$ & \{0,1\} & Binary variable of acoustic cladding. Only used for fans and VFCs.
 \\\\
 parameter  & domain & description \\ 
 \hline
 $\lwx{\mathrm{in}}_{f,s}$ & $[\underline{\lw},\overline{\lw}]$ & octave sound power level input with octave band $f\in\mathcal{F}$ in scenario $s\in\mathcal{S}$ \\
 $\lwx{\mathrm{out}}_{f,s}$ & $[\underline{\lw},\overline{\lw}]$ & octave sound power level output with octave band $f\in\mathcal{F}$ in scenario $s\in\mathcal{S}$\\
 $z_{f,s}$ & $\{0,1\}$ & Binary variable for the maximisation of octave sound power levels, see \cref{sec:lvl_add}. With octave band $f\in\mathcal{F}$ in scenario $s\in\mathcal{S}$\\
 $\lwx{\mathrm{inc}}_{f,s}$ & $[\underline{\lw},\overline{\lw}]$ & increase of octave sound power level due to component with octave band $f\in\mathcal{F}$ in scenario $s\in\mathcal{S}$\\
 $\lwx{\mathrm{max}}_{f,s}$ & $[\underline{\lw},\overline{\lw}]$ & maximum octave sound power level of damped input and flow noise with octave band $f\in\mathcal{F}$ in scenario $s\in\mathcal{S}$\\
 $\Delta \lw_{f,s}$ & $[\underline{\lw},\overline{\lw}]$ & absolute difference of octave sound power level damped input and flow noise with octave band $f\in\mathcal{F}$ in scenario $s\in\mathcal{S}$
\end{tabular}
\end{table}

%% file: Results.tex
 \newpage
\section{Results and Discussion}
\label{sec:results}

This section demonstrates and discusses the application of the mathematical optimisation model to a real-world example. It examines the trade-off between life-cycle costs and noise limits, providing a transparent analysis, and investigates the differences between sequential and holistic optimisation approaches. The section concludes with a discussion of the study’s limitations.

\subsection{Case study}
\label{sec:case_study}

The case study uses an existing multi-purpose building consisting of two supply and three exhaust air systems which are all central systems. For the case study, only the main air supply system is investigated. The separation from supply air and exhaust air system is possible, as discussed in \cref{sec:method_overview}. The system uses $\mathrm{CO_2}$ control systems. Accordingly it makes sense to consider different load scenarios.

\begin{figure}[htbp]
\centering
\includegraphics[width=0.8\textwidth]{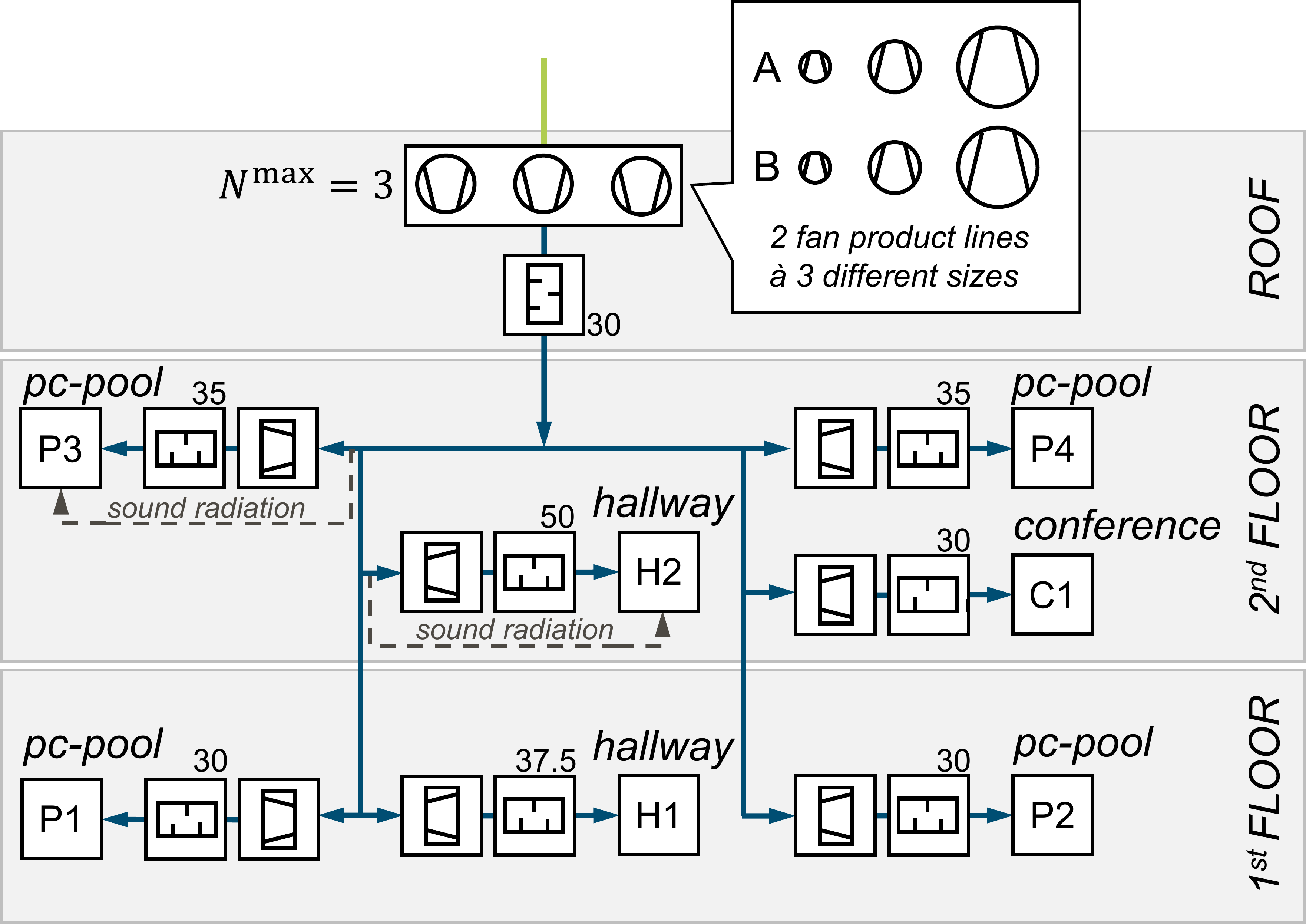}
\caption{schematic representation of the case study building. The ventilation system under consideration ventilates the major rooms in the building's first and second floor. It supplies seven rooms of which most are pc-pools. The value next to silencers indicate below which noise limits in $\si{\decibel}$ the silencer was bought.} \label{fig:case_study_topology}
\end{figure}

As a first step, the load cases are obtained in form of volume flow requirements into the rooms. Then, the respective pressure losses in each duct section are calculated using the CAD software REVIT~\cite{revit}. The pressure losses of VFCs and silencers are not included as these are part of the optimisation. The changes in sound power level for each component are calculated according to VDI 2081. The volume flow requirements and the respective pressure losses as well as the noise limits for each room are publicly accessible; refer to \cref{sec:data_availability} for details. The fixed elements in the duct network are combined if applicable, see \cref{sec:fixed_comp}. This leads to an optimisation system with 38 components of which 19 are fixed. The network without fixed components is shown in in \cref{fig:case_study_topology}. The sound pressure limits are taken from the building's planning data. They differ between $\qtyrange{40}{45}{\decibel}$. For two rooms, namely $P3$ and $H2$, the sound radiation through the duct is also taken into account for the sound pressure limits. This is necessary because these rooms are the most susceptible to noise.
The changes in sound power level throughout the network are verified - where possible - using the EXCEL calculation tool provided in VDI standard 2081.
The fan station can have up to three fans ($N^\mathrm{max}=3$), which can be chosen from a set of two product lines à three different diameters where each fan can be chosen twice, leading to twelve different fans. VFCs in front of each room are set to be purchased. This ensures that solutions remain viable not only for the load cases under consideration but also for other potential load cases. The total annuity costs of VFCs of €23,000, are not factored into the life-cycle costs.

The mathematical models are implemented in Python using the PYOMO~\cite{hart2011pyomo} modeling language and then solved using Gurobi 11.0.0~\cite{gurobi}. Gurobi solves the problem using a spatial branch-and-bound algorithm. The global optimality of the solution is guaranteed by a primal and a dual bound. The calculations were performed on a Lenovo P14s laptop, which features a 64-bit Windows 11 operating system, 32 GB of RAM, and an AMD Ryzen 7 PRO 5850U processor with 16 cores and a maximum clock speed of 4.4 GHz.

\subsection{Optimisation results}
\label{sec:opt_res}
For comparing the airflow-only with the holistic optimisation solution, both problems are solved independently. The problem sizes, including the number of constraints and variables, as well as the computation time and resulting life-cycle costs for both the airflow-only and coupled systems, are presented in \cref{tab:res}. The increase in constraints and variables is more than tenfold, it mostly stems from introducing the eight sound power levels per component and load case and its additional binary variables.
This increase also results in increased computation times. While the airflow-only approach solves within $\SI{1}{\second}$, the coupled approach takes $\SI{24}{\second}$ to reach global optimality. This increase in computation time is a consequence of the substantial rise in system complexity. Nevertheless, a solution time of $\SI{24}{\second}$ remains remarkably efficient given the scale of the coupled approach.
\begin{table}[htbp]
    \centering
    \caption{Problem size, optimisation results and computation time.}
    \label{tab:res}
    \begin{tabular}{  p{3.5cm} p{3cm} p{3cm}}
      & airflow-only & airflow + acoustics\\ 
      \hline
     \# constraints & 792 & 14,523  \\  
     \# continuous variables & 414 & 5,022 \\
     \# integer variables & 104 & 1,349 \\
     \# binary variables & 80 & 1,325\\
     computation time & 2~s & 24~s\\
     min life-cycle costs & 8,576~€ & 10,010~€
    \end{tabular}
\end{table}
The life-cycle costs increase by 17~\% due to integration of the acoustics. The resulting topology is shown in \cref{app:case_study_solution}. In both optimizations, the same single fan is used across all load cases. Notably, silencers in front of most rooms are not necessary to reach the noise limits. Only in the hallways silencers are needed due to the significant throttling by the VFCs and the presence of loud air outlets. In the pc-pools large induction ceiling diffusers are used which are comparably quite. The result is remarkable as in the traditional planning procedure silencers are placed, often without thoroughly validating their necessity. These results suggest that substantial investment costs can be saved by calculating the actual need for silencers. Consequently, operating costs remain unchanged, with the increase in life-cycle costs attributed solely to higher investment costs.

\begin{figure}[htpb]
     \centering
     \begin{subfigure}[b]{0.4\textwidth}
        \centering
    	\includegraphics[width=\textwidth]{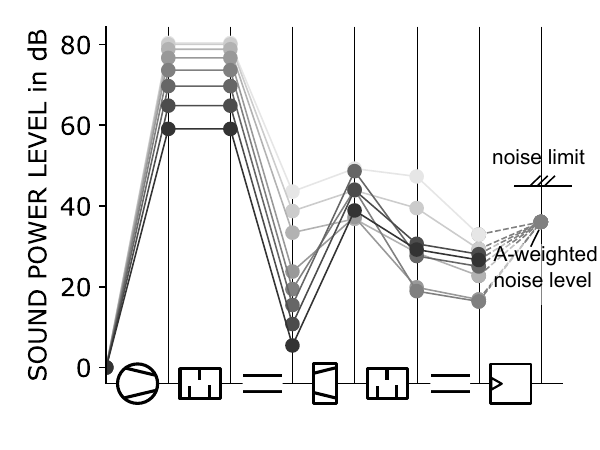}
    	\caption{Room: H2}
    	\label{fig:ospl_h2}
     \end{subfigure}
     \hfill
     \begin{subfigure}[b]{0.4\textwidth}
         \centering
         \includegraphics[width=\textwidth]{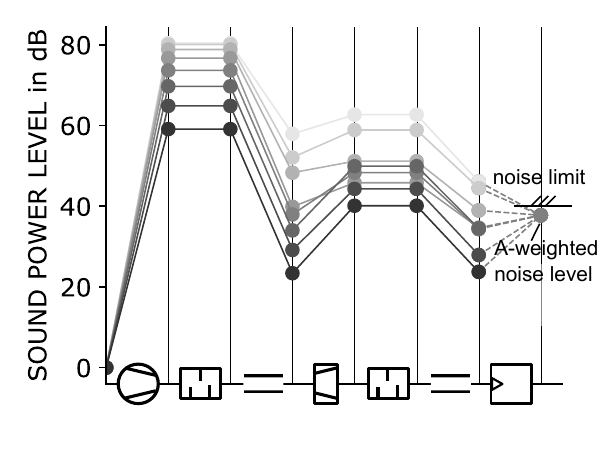}
         \caption{Room: P3}
         \label{fig:ospl_p3}
     \end{subfigure}
     \hfill
     \begin{subfigure}[t]{0.13\textwidth}
         \centering
         \vspace{-35ex} 
         \includegraphics[width=\textwidth]{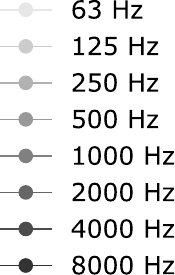}
     \end{subfigure}
     \caption{octave sound power level changes and the A-weighted room noise level for two different branches in the network for the maximum load case. The icons on the x-axis indicate which element causes the change. The horizontal duct represents fixed elements. The A-weighted noise level is always below the noise limits. Even with the active VFC in front of room P3, the noise level remains at an acceptable level throughout all scenarios, thus no silencer is purchased.}
    \label{fig:ospl_results}
\end{figure}

For an in-depth analysis, in \cref{fig:ospl_results} the change in octave sound power levels is shown for two different paths from the central fan station to the respective room. For the remaining branches, the results are given in \cref{app:ospl_changes_all_rooms}. In all cases, the A-weighted noise level is well below the noise limit. For room H2 a small silencer is necessary that dampens the high frequency noise induced by the VFC. For room P3, the duct network dampens the noise similar to H2, yet the air outlet keeps the noise at an acceptable level making a silencer unnecessary.

\subsection{Minimal Life-Cycle Costs vs. Minimal Noise}
\label{sec:lcc_vs_ac}
Minimising the life-cycle costs while keeping the noise at an acceptable level is an important step. Yet, the presented methodology even allows making the trade-off between minimal life-cycle costs and minimal noise transparent. Therefore, the approach outlined in \cref{sec:solving_algorithm} is used: First, the system without acoustics is optimised. Then, the holistic optimisation problem is solved with iteratively decreasing sound pressure levels until no feasible solution can be found. For simplicity, noise limits for all rooms are set to be identical.

\begin{figure}[htbp]
    \centering
    \includegraphics[width=0.6\textwidth]{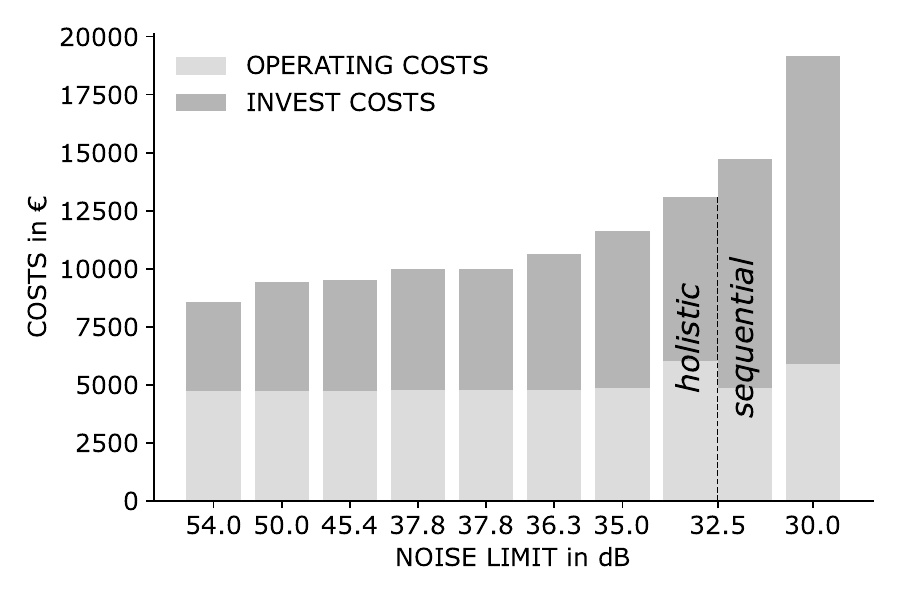}
    \caption{life-cycle costs of the systems with differing sound pressure level limits calculated with the coupled approach. For the $\SI{32.5}{\decibel}$ noise limit solution, the sequential approach yields a different solution, see \cref{sec:sequential}. Here the life-cycle costs are 12~\% higher than in the holistic approach. The data for this figure is publicly accessible; refer to \cref{sec:data_availability} for details.}
    \label{fig:costs_split}
\end{figure}

In \cref{fig:costs_split}, the operating and invest costs are shown for the different noise limits. The noise limits set for the optimisation were often not fully reached, as the system selected a topology that resulted in actual noise levels being even lower. In such cases, the highest resulting noise level in the room was adopted as the new noise limit. Consequently, the noise limits shown in the figure are no longer integers. The resulting system life-cycle costs more than double from the airflow-only solution to the one with minimal noise limit. As expected, with decreasing sound pressure limits, more silencers are purchased and more dampening is induced by the silencers as their length and number of splitter elements increases. See values next to silencer icons in \cref{fig:case_study_topology} indicating from which noise limit on they are purchased. The silencer behind the central fan station is only chosen for the lowest noise level. This further emphasises the importance of thoroughly calculating the necessity of silencers. The operating costs do not change above a sound power level of $\SI{32.5}{\decibel}$. The increase in invest costs from $\SI{32.5}{\decibel}$ to $\SI{30}{\decibel}$ stems mostly from the central silencer, which is relatively costly because of its large size.

The computation times for most noise limits are comparable to those discussed in \cref{sec:opt_res}. However, the lowest noise limit of \(\SI{30}{\decibel}\) results in a significant increase in computation time, reaching \(\SI{14}{\hour}\). Although the exact reasons for the increased computation time in MINLP solvers are difficult to pinpoint, it typically indicates the presence of multiple nearly equivalent optimal decisions. This is further evidenced by the solver's gap of 1~\% after just \(\SI{20}{\min}\), highlighting the challenge of identifying the optimal solution in such cases. The detailed computation times are given in \cref{app:comp_time}.

\begin{figure}[htbp]
\centering
\includegraphics[width=0.8\textwidth]{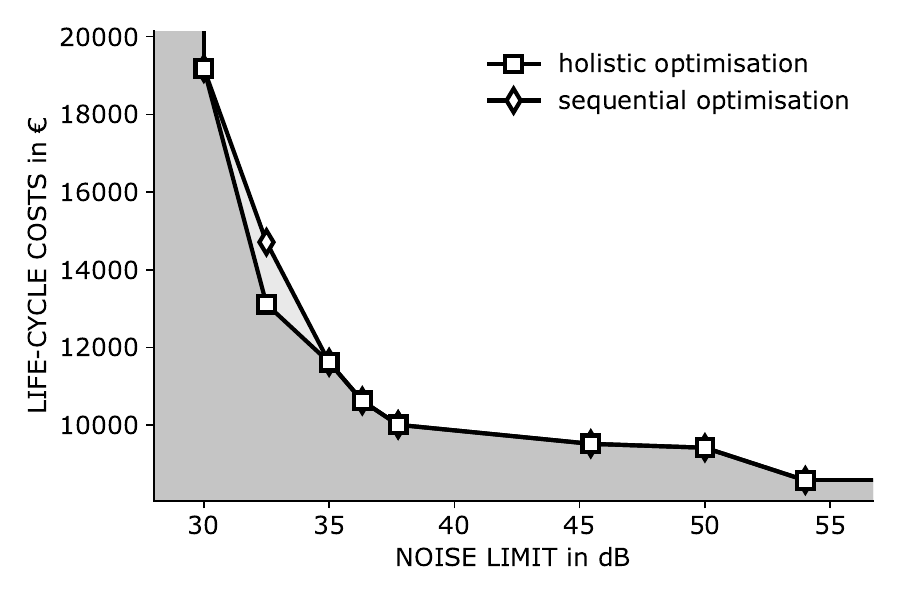}
\caption{
Pareto fronts demonstrating the trade-off between minimising life-cycle costs and noise levels for both holistic and sequential optimization approaches. All displayed solutions are optimal with respect to life-cycle costs while adhering to a predefined noise limit. The solution with the lowest life-cycle cost is derived from optimization focused exclusively on airflow, with its noise level calculated during postprocessing. All other solutions are obtained by holistic optimisation of airflow and acoustics. Markers representing sequential optimisation closely align with those of holistic optimisation, making them nearly indistinguishable except at the $\SI{32.5}{\decibel}$ noise limit. Detailed data for this plot is publicly available; see \cref{sec:data_availability} for access information.} \label{fig:pareto_front}
\end{figure}

To make the multi-criterial decision-space more transparent for decision makers the results are visualised using a Pareto-front, see \cref{fig:pareto_front}. This contains the optimal solutions, showing the conflict between sound pressure level limits and life-cycle costs. 
While the solutions for noise limits of $\SI{32.5}{\decibel} - \SI{37.5}{\decibel}$ show a good trade-off between life-cycle costs and noise limit, reducing either the noise limit or the life-cycle costs further, leads to an enormous increase in the other objective's value. Concretely, the minimal noise limit case of $\SI{30}{\decibel}$ comes with a $46~\%$ increase in life-cycle cost compared to the $\SI{32.5}{\decibel}$ noise limit solution. On the other hand, choosing a noise limit of $\SI{46}{\decibel}$ compared to $\SI{37.5}{\decibel}$ reduces the life-cycle costs by $5~\%$, a limit of $\SI{50}{\decibel}$ only brings $1~\%$ further improvement in life-cycle costs.
Regarding the ventilation system topology, all solutions chose the same fan configuration: a single operating fan. For all but but the solution for a noise limit of $\SI{32.5}{\decibel}$, the exact same fan is chosen. For the noise limit of $\SI{32.5}{\decibel}$ a cheaper but less efficient fan is chosen. The reason for this choice is analysed in \cref{sec:sequential}.
It is remarkable that noise limits have virtually no impact on fan selection. This outcome arises because, for most noise limits, silencers are not installed in branches with the highest total pressure loss. As a result, the maximum pressure loss that the fan station must overcome remains unaffected. This observation suggests a weak correlation between energy efficiency and acoustics.

\subsection{Holistic vs. Sequential Optimisation}
\label{sec:sequential}

The Pareto front described above is obtained using holistic optimisation, but the proposed method also supports sequential optimisation for the coupled airflow and acoustic problem. In sequential optimisation, the process is divided into two steps. First, only airflow is considered, determining the selection, placement, and operation of fans and VFCs to minimise life-cycle costs. In the second step, airflow and acoustics are optimised together, but the fan and VFC purchase decisions are fixed based on the first step. While holistic optimisation guarantees a globally optimal solution by solving a coupled physical model, sequential optimisation offers the advantage of simpler models, requiring fewer decisions per optimisation step. This results in faster solving times but may lead to higher life-cycle costs or infeasible solutions. Solving the sequential optimisation problem produces globally optimal solutions with respect to life-cycle costs in each of the two steps. This represents a significant improvement even over traditional sequential planning procedures.

The question of topological differences between the sequential and holistic solutions is implicitly addressed in \cref{sec:opt_res}. For the airflow-only solution, a single fan is purchased and operating in all load cases. When coupling airflow and acoustics, the topology remains unchanged for all cases except for the \(\SI{32.5}{\decibel}\) noise limit. Thus, the acoustics have no influence on the fan station topology except for this one noise limit. Consequently, the sequential solution aligns with the holistic optimisation for all but the \(\SI{32.5}{\decibel}\) case. At this noise limit, the holistic approach involves purchasing a different fan than the sequential approach, resulting in life-cycle costs that are 12\% lower than those of the holistic optimisation, as shown in \cref{fig:costs_split}. This significant cost decrease is primarily due to topological differences. Unlike the sequential optimisation, the holistic approach eliminates the need for a central silencer, resulting in considerably lower investment costs. While this reduction in investment costs comes at the expense of energy efficiency, the savings are sufficient to offset the higher energy costs. Consequently, for this noise limit, the holistic optimisation proves more cost-effective than the sequential approach.
\newline
These computation times demonstrate that, while the holistic optimisation approach generally requires more computational effort: It averages $\SI{27}{\s}$ over all noise limits but the $\SI{30}{\decibel}$ solution which is significantly higher ($\SI{14}{\hour}$). The times remain reasonable given the complexity of the problem. The sequential approach, with average computation times of $\SI{2}{\s}$ for airflow-only and $\SI{4}{\s}$ for the coupled optimisation, is computationally more efficient but does not fully capture the interplay between airflow and noise constraints. This highlights the trade-off between computational expense and the level of optimisation achieved.

\subsection{Limitations}

This study makes several assumptions regarding the acoustic modeling of noise sources within the building, which introduce limitations to the analysis. One key assumption is the focus on noise generated by the supply air system while neglecting the potential contribution of noise from the exhaust air system. This simplification does not account for scenarios where the noise levels of the supply and exhaust systems are comparable. In such cases, the cumulative noise level could increase by up to \(\SI{3}{\decibel}\). However, this limitation is considered to have a minimal impact on the overall results, as the resulting noise in the rooms are well below the limits for all but the very low noise limit conditions. Future investigations could address this issue by optimising both supply and exhaust air systems and incorporating their combined sound pressure levels into the analysis.

The model equations, mostly linear in the respective variables, are obtained by relying on model equations from the literature in combination with data from manufacturers. Where literature model equations were absent, model equations were obtained solely by approximating manufacturer data. This raises the question of how viable these models are for other manufacturer's components. Here, further research could explore the viability of component models in greater depth.

Another assumption pertains to the modeling of the fan station's sound power levels. The current approach simplifies the calculation by considering the sound power level of the fan station as equal to the maximum sound power level of any individual fan, as formulated in \cref{eq:fan_spl2}. This assumption neglects the cumulative effect when multiple fans with similar sound power levels operate simultaneously, which would result in an increased noise level of several decibels. However, this underestimation is justified in the context of this study, as all solutions involve the purchase of a single fan. Consequently, the assumption does not affect the global optimality of the results. If scenarios with multiple fans were to be considered in future research, a more accurate approach incorporating cumulative sound power levels —similar to the level addition used for room acoustics, c.f. \cref{eq:seven_level_additions} — would be advisable.

In addition to the method-related limitations discussed, it is important to emphasise that the accuracy of the octave sound power level approximation in the network is inherently tied to the underlying VDI 2081 standard. As this standard includes multiple simplifications, it likely accounts for the largest deviations when compared to a real system. Addressing these deviations requires improving the availability of data, which could enable the use of more advanced approaches, such as an acoustic four-pole method as discussed in \cref{sec:acoustic_modeling}.

%% file: Discussion.tex
\section{Conclusion}
\label{sec:conclusion}

In this study, an innovative methodology for designing ventilation systems via mathematical optimisation is introduced. In order to prioritise both life-cycle costs - including energy costs - and acoustic comfort a 2-stage stochastic MINLP is derived, advancing the current body of research in several key areas. While existing research focuses on efficiently ensuring indoor air quality, the presented methodology additionally considers acoustics as an integral part of ventilation systems design. To the best of our knowledge, this is the first paper integrating the acoustics in the algorithmic design of a ventilation system.

In order to incorporate the coupling between airflow and acoustics, VFCs, silencers and fixed components are modelled. New models are introduced for each of the components that offer simple yet accurate physical model equations in the realm of ventilation system airflow and acoustics.

Through a case study, the approach's viability in balancing life-cycle costs and acoustic comfort was illustrated, supporting building owners to make informed decisions regarding the cost of acoustic comfort. The presented holistic optimisation problem is significantly larger than problems focusing solely on airflow - more than tenfold in size. However, major reductions of model size and of non-linearity can be achieved by exploiting the system structure. Further improvements can be achieved by using an efficient solving algorithm.
The resulting MINLP is solvable within a reasonable computational timeframe. The results are globally optimal in terms of life-cycle costs while assuring noise limits according to the standard VDI 2081. Remarkably, compared to the real case study building, the proposed methodology suggests installing significantly fewer silencers. This highlights the importance of carefully evaluating where silencers are necessary and where they can be avoided. Such an investigation is challenging for human planners due to the complexity of the decision-making process, as discussed in \cref{sec:intro}.

The connection between life-cycle costs and noise limits was made transparent using a Pareto front. It becomes evident that while multiple solutions within the range of \(\SI{32.5}{\decibel} - \SI{37.5}{\decibel}\) provide reasonable trade-offs, further reducing one objective significantly worsens the other. This insight enables decision-makers to make transparent and informed choices. Furthermore, it was demonstrated that the connection between energy efficiency and acoustics is minimal in most cases. Stricter noise limits had little impact on the branches with the highest pressure loss and, as a result, did not significantly increase the pressure rise required by the fan.
This observation aligns with the insight gained when comparing traditional sequential planning with the holistic optimisation approach. For the sequential planning approach, two optimisation problems are solved with regard to minimal life-cycle costs. Thus, the subproblems of the sequential planning are each also solved globally optimal - this is a major improvement even for the traditional planning procedure. Findings suggest that although the holistic method is more time-consuming, it does not always yield superior outcomes. For most noise levels considered, both approaches lead to identical results. However, at a noise limit of $\SI{32.5}{\dB}$, the holistic approach modifies the system topology and reduces the life-cycle costs by 12~\%. To do so, a less efficient fan is accepted as a trade-off to save on overall life-cycle expenses. This shows that holistic optimisation indeed allows for a tremendous decrease in life-cycle costs. To transparently illustrate the relationship between energy consumption, investment costs, and acoustics, future research could optimise for not two but three conflicting objectives is essential. The resulting Pareto front in 3D space would effectively elucidate these interdependencies.

In future work, the case study should be extended to take into account more complex buildings, possibly allowing distributed fans in the central duct network. Then, potentially the holistic optimisation approach unfolds its true potential.



%% file: Appendix.tex
\section{Planning Procedures in Different Countries}
\label{app:planning_procedures}
\begin{figure*}[htb]
	\centering
		\includegraphics[width=0.95\linewidth]{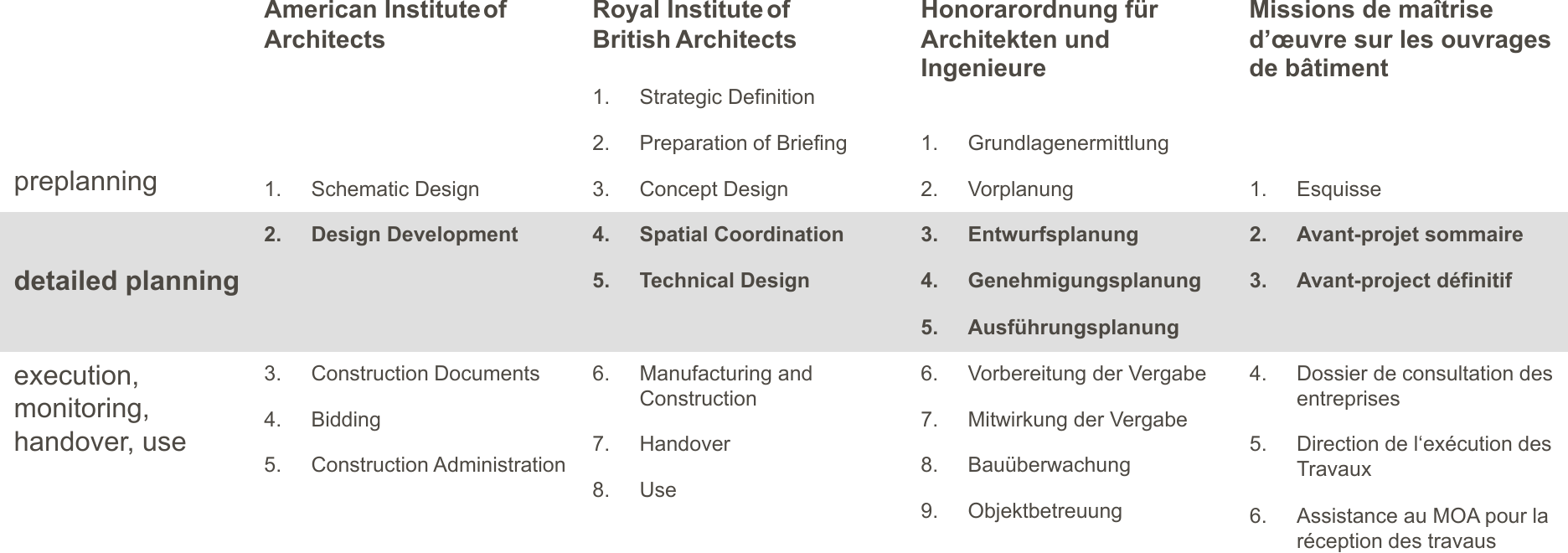}
	\caption{overview of planning procedures in different countries.}
	\label{fig:hoai_einordnung}
\end{figure*}

The design of a ventilation system is part of a highly complex process that is necessarily divided into many subtasks that are to be performed in parallel, sequentially and also iteratively. In many countries, regulatory frameworks define the planning process such as the \emph{Plan of Work} in the UK by the Royal Institute of British Architects~\cite{PlanOfWork}, 
such as the \emph{Plan of Work} in the UK by the Royal Institute of British Architects~\cite{PlanOfWork}, Germany's \emph{Honorarordnung für Architekten und Ingenieure}~\cite{HOAI}, the American Institute of Architects' \emph{standard form} in the USA~\cite{AIAStandardForm2017}, and France's \emph{Missions de maîtrise d’œuvre sur les ouvrages de bâtiment}~\cite{MissionsMaitriseOeuvre}. Despite varying in rigor, detail, and scope, these regulations generally encompass three main phases: preplanning, detailed planning, and construction oversight, see \cref{fig:hoai_einordnung}. Therefore, focusing solely on the detailed planning phase, while excluding preplanning, is an approach valid for all stated countries.

\section{Annuity Cost Calculation}
\label{app:annuity}
The annuity costs are calculated based on the guidelines in \cite{VDI2067}. Each component has a depreciation period, $T^{\mathrm{dep},com}$, as specified in VDI 2067. For VFCs no specific data is provided, thus fan depreciation periods are used. If a component's depreciation period is shorter than the planned operational life of the ventilation system, replacement purchases are required. The total investment for the planned operational period is calculated as:

\begin{align}
    I^\mathrm{tot} = \sum_{i=1}^{T^\mathrm{div} - \delta} \left( \frac{{R^\mathrm{M,S}}}{Z} \right)^{T^{\mathrm{dep},com} \cdot i},
\end{align}
where $T^\mathrm{div} = \left\lfloor \frac{T^\mathrm{use}}{T^{\mathrm{dep},com}} \right\rfloor$ and $\delta$ equals 1 if the condition $\{T^\mathrm{use} \mod T^{\mathrm{dep},com} = 0\}$ holds and 0 otherwise. For any (replaced) component whose depreciation period exceeds the planned operation years, the residual value is calculated as follows:

\begin{align}
    R^\mathrm{W} = \frac{T^\mathrm{use} \mod T^{\mathrm{dep},com}}{T^{\mathrm{dep},com}} \frac{{R^\mathrm{M,S}}^{(T^{\mathrm{dep},com} T^\mathrm{div})}}{Z^T},
\end{align}
where $\mod$ is the modulo operator.

The maintenance $F^{c,\mathrm{M}}$ and service factors are considered proportional to the invest costs and component-dependent, $F^{c,\mathrm{S}}$. Finally, the annuity investment cost factor $A^\mathrm{F}$ is determined by:

\begin{align}
    A^{\mathrm{F},com} = A \left( 1 + B^\mathrm{S,M}(F^{\mathrm{S},com} + F^{\mathrm{M},com})\right)\left( 1 + I^\mathrm{tot} - R^\mathrm{W} \right).
\end{align}

This approach allows a comprehensive calculation of annuity costs, taking into account depreciation, replacement needs, and residual values.

\section{Case Study solution}
\label{app:case_study_solution}
\begin{figure}[htbp]
    \centering
    \includegraphics[width=0.8\textwidth]{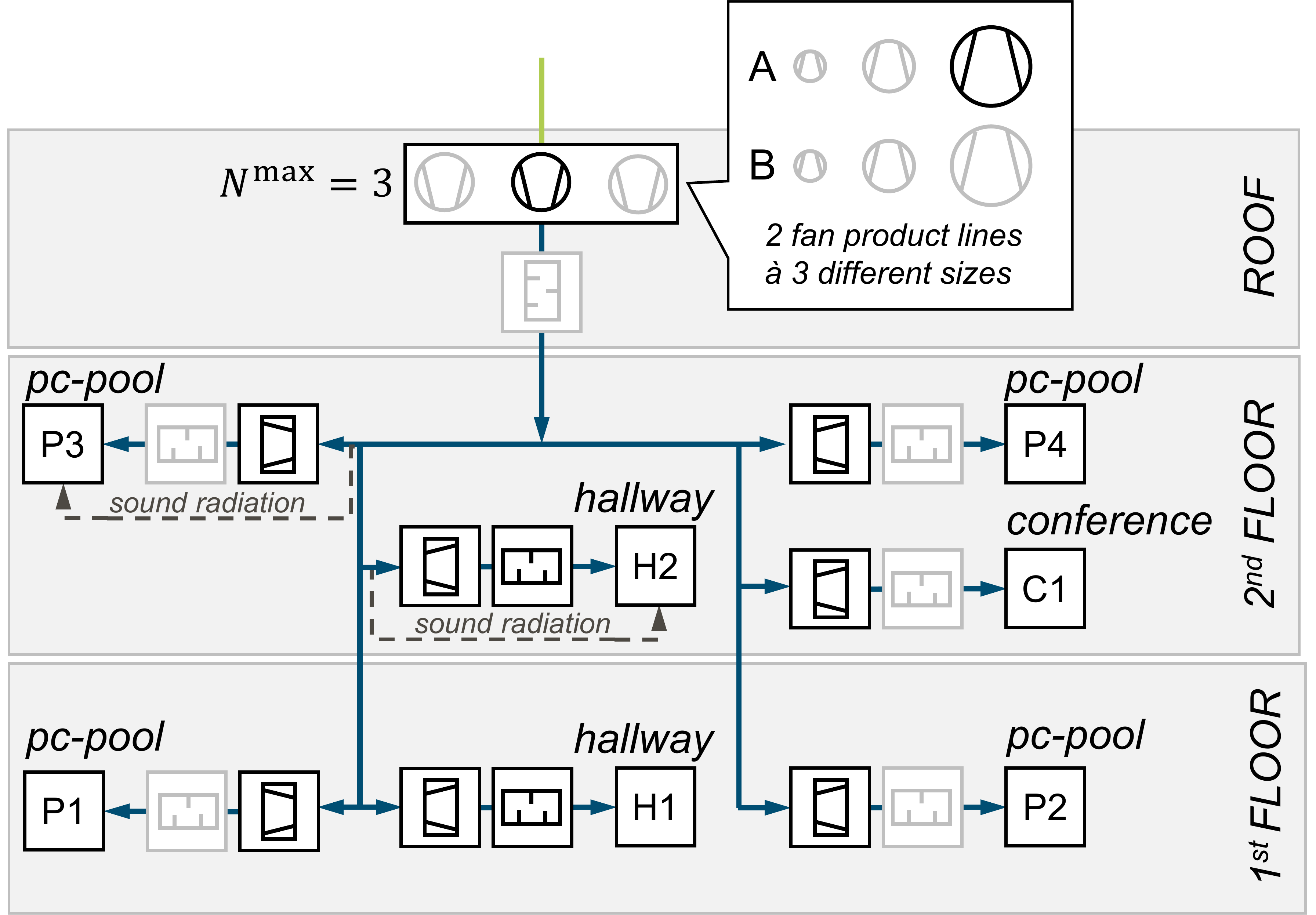}
    \caption{solution of the case study building with noise limits in the rooms. Not purchased components are greyed out. Only one fan is purchased in the central fan station. Many silencers in front of the rooms do not have to be purchased. Only two silencers are needed in front of the hallways H1, H2.} \label{fig:case_study_topology_real}
\end{figure}

The solution topology for the case study from \cref{sec:opt_res} is shown in \cref{fig:case_study_topology_real}.

\section{Octave Sound Power Level Changes for all Branches}
\label{app:ospl_changes_all_rooms}
\begin{figure}[htbp]
    \centering
    \includegraphics[width=0.8\textwidth]{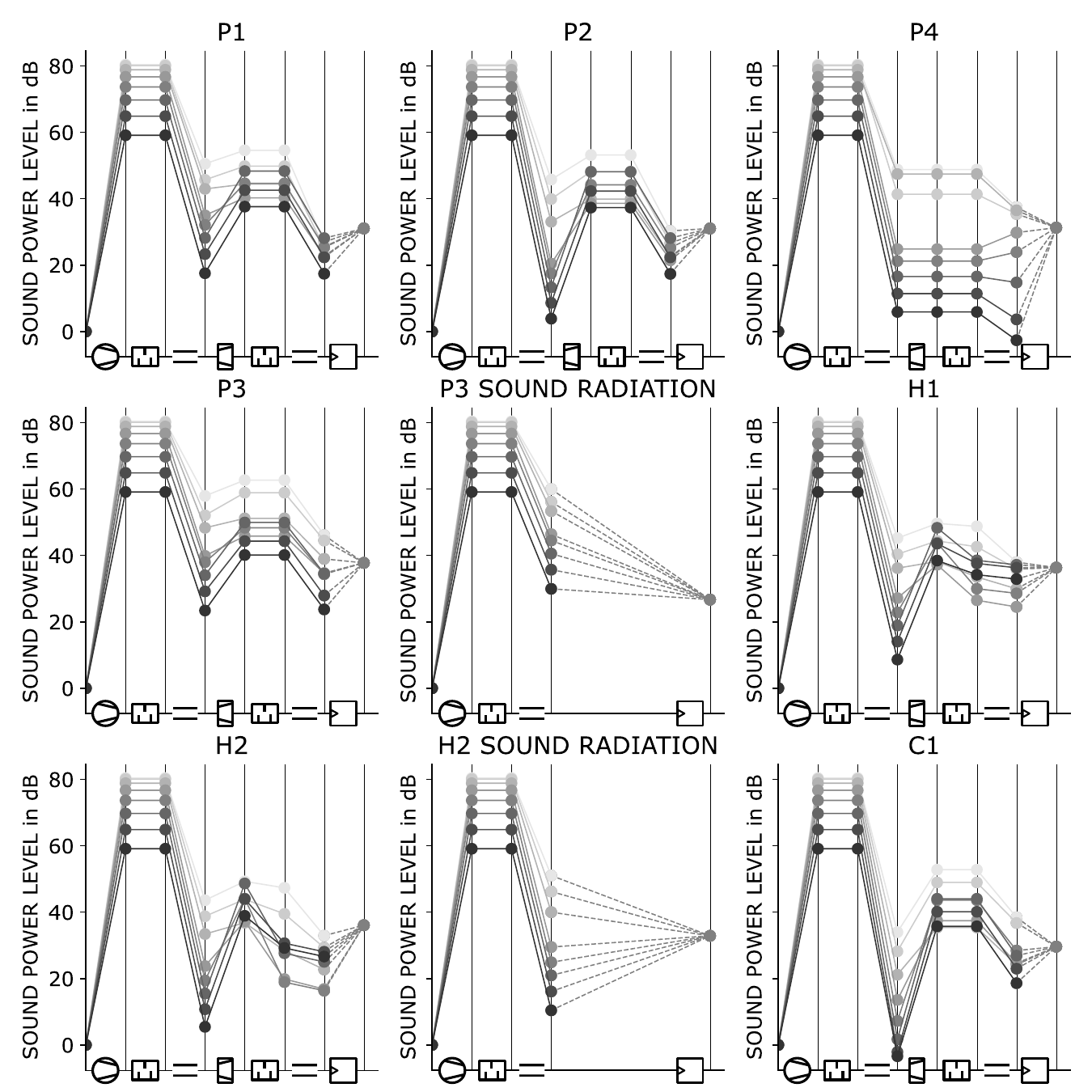}
    \caption{octave sound power level changes and the A-weighted room noise level for all branches in the network for the maximum load case. The icons on the x-axis indicate which element causes the change. The horizontal duct represents fixed elements.} \label{fig:ospl_all}
\end{figure}

In addition to \cref{fig:ospl_results}, the octave sound power level changes for all branches are given in \cref{fig:ospl_all}.

\section{Computation Times}
\label{app:comp_time}
\begin{figure}[htbp]
    \centering
    \includegraphics[width=0.6\textwidth]{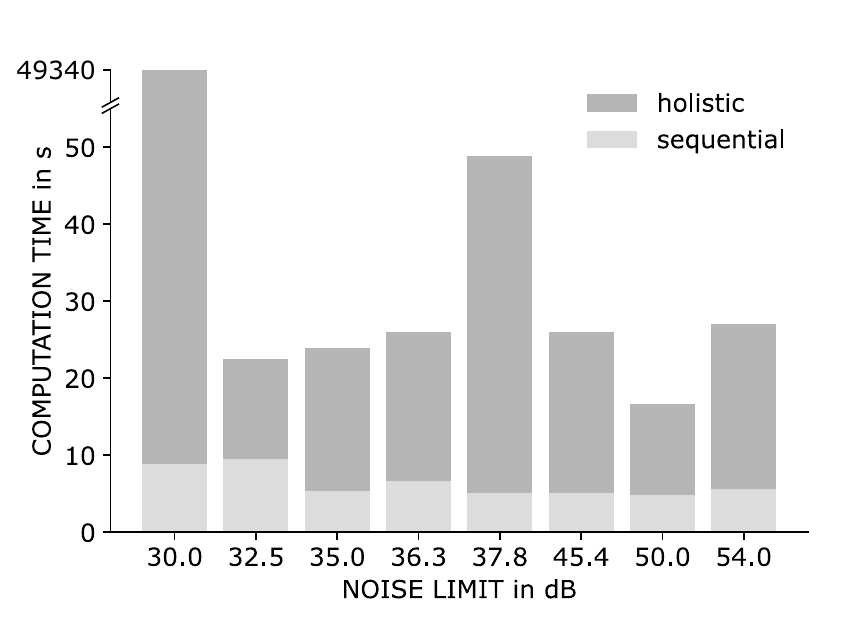}
    \caption{computation times for the holistic and sequential optimisation for all noise limits.} \label{fig:comp_time}
\end{figure}

The computation times for all optimisations are shown in \cref{fig:comp_time}.

\section{Data availability}
\label{sec:data_availability}
The load cases used within the optimisations as well as the data behind \cref{fig:ospl_results,fig:costs_split,fig:pareto_front,fig:ospl_all,fig:comp_time} are openly available on the data repository TUdatalib \url{https://tudatalib.ulb.tu-darmstadt.de/handle/tudatalib/4232.2}.

%% file: main.bbl
\begin{thebibliography}{10}
\expandafter\ifx\csname url\endcsname\relax
  \def\url#1{\texttt{#1}}\fi
\expandafter\ifx\csname urlprefix\endcsname\relax\def\urlprefix{URL }\fi
\expandafter\ifx\csname href\endcsname\relax
  \def\href#1#2{#2} \def\path#1{#1}\fi

\bibitem{FederalMinistryforEconomicAffairsandEnergy.2015}
{Federal Ministry for Economic Affairs and Energy}, Energy efficiency strategy
  for buildings: Methods for achieving a virtually climate-neutral building
  stock (2015).

\bibitem{carrie2017impact}
F.~R. Carri{\'e}, V.~Leprince, M.~Kapsalaki, Impact of energy policies on
  building and ductwork airtightness, in: Building Simulation, Vol.~9,
  Springer, 2017, pp. 359--398.

\bibitem{en12081414}
P.~Šujanová, M.~Rychtáriková, T.~Sotto~Mayor, A.~Hyder,
  \href{https://www.mdpi.com/1996-1073/12/8/1414}{A healthy, energy-efficient
  and comfortable indoor environment, a review}, Energies 12~(8) (2019).
\newblock \href {https://doi.org/10.3390/en12081414}
  {\path{doi:10.3390/en12081414}}.
\newline\urlprefix\url{https://www.mdpi.com/1996-1073/12/8/1414}

\bibitem{VDI3803-1}
{Verein Deutscher Ingenieure}, {VDI 3803 - 1}: Air-conditioning: Structural and
  technical principles, central air conditioning systems (2018).

\bibitem{en16041853}
L.~Amanowicz, K.~Ratajczak, E.~Dudkiewicz,
  \href{https://www.mdpi.com/1996-1073/16/4/1853}{Recent advancements in
  ventilation systems used to decrease energy consumption in buildings -
  literature review}, Energies 16~(4) (2023).
\newblock \href {https://doi.org/10.3390/en16041853}
  {\path{doi:10.3390/en16041853}}.
\newline\urlprefix\url{https://www.mdpi.com/1996-1073/16/4/1853}

\bibitem{DeutscheEnergieAgentur.2018}
{Deutsche Energie-Agentur}, Ratgeber: Lufttechnik f{\"u}r industrie und gewerbe
  (2018).

\bibitem{stanford2019analysis}
H.~W. Stanford~III, A.~F. Spach, Analysis and design of heating, ventilating,
  and air-conditioning systems, CRC Press, 2019.

\bibitem{DeutscheEnergieAgenturGmbH.2010}
{Deutsche Energie-Agentur GmbH}, Ratgeber: Lufttechnik f{\"u}r industrie und
  gewerbe (2010).

\bibitem{mckane2003improving}
A.~T. Mckane, Improving compressed air system performance: A sourcebook for
  industry, U.S. Department of Energy Energy Efficiency and Renewable Energy,
  2003.

\bibitem{DIN16798-3}
{Deutsches Institut für Normung e.V.}, {DIN EN 16798 - 3}: Energy performance
  of buildings – ventilation for buildings (2017).

\bibitem{Alaraj.2022}
M.~Ala'raj, M.~Radi, M.~F. Abbod, M.~Majdalawieh, M.~Parodi, Data-driven based
  hvac optimisation approaches: A systematic literature review, Journal of
  Building Engineering 46 (2022) 103678.
\newblock \href {https://doi.org/10.1016/j.jobe.2021.103678}
  {\path{doi:10.1016/j.jobe.2021.103678}}.

\bibitem{sha2019overview}
H.~Sha, P.~Xu, Z.~Yang, Y.~Chen, J.~Tang, Overview of computational
  intelligence for building energy system design, Renewable and Sustainable
  Energy Reviews 108 (2019) 76--90.

\bibitem{ahmad2016computational}
M.~W. Ahmad, M.~Mourshed, B.~Yuce, Y.~Rezgui, Computational intelligence
  techniques for hvac systems: A review, in: Building Simulation, Vol.~9,
  Springer, 2016, pp. 359--398.

\bibitem{acuna2014review}
E.~I. Acu{\~n}a, I.~S. Lowndes, A review of primary mine ventilation system
  optimization, Interfaces 44~(2) (2014) 163--175.

\bibitem{schanzle2015good}
C.~Sch{\"a}nzle, L.~C. Altherr, T.~Ederer, U.~Lorenz, P.~F. Pelz, As good as it
  can be-ventilation system design by a combined scaling and discrete
  optimization method, tuprints (2015).

\bibitem{muller2023planning}
T.~M. M{\"u}ller, M.~Sachs, J.~H. Breuer, P.~F. Pelz, Planning of distributed
  ventilation systems for energy-efficient buildings by discrete optimisation,
  Journal of Building Engineering 68 (2023) 106205.

\bibitem{eilemann1999practical}
A.~Eilemann, Practical noise and vibration optimization of hvac systems, Tech.
  rep., SAE Technical Paper (1999).

\bibitem{luzzato2019aero}
C.~M. Luzzato, J.~Biermann, R.~Fouque, The aero-acoustic design and
  optimization of a ground transportation hvac system using lattice boltzmann
  methods, in: NAFEMS World Congress, 2019, p.~1.

\bibitem{Ferrara.2021}
M.~Ferrara, J.~C. Vall{\'e}e, L.~Shtrepi, A.~Astolfi, E.~Fabrizio, A thermal
  and acoustic co-simulation method for the multi-domain optimization of nearly
  zero energy buildings, Journal of Building Engineering 40 (2021) 102699.
\newblock \href {https://doi.org/10.1016/j.jobe.2021.102699}
  {\path{doi:10.1016/j.jobe.2021.102699}}.

\bibitem{DIN.18599-10}
{Deutsches Institut für Normung e.V.}, {DIN18599 - 10}: Energetische bewertung
  von geb{\"a}uden: Berechnung des nutz-, end- und prim{\"a}renergiebedarfs
  f{\"u}r heizung, k{\"u}hlung, l{\"u}ftung, trinkwarmwasser und beleuchtung -
  teil 10: Nutzungsrandbedingungen, klimadaten (2016).

\bibitem{DIN16798}
{Deutsches Institut f{\"u}r Normung e. V.}, {DIN EN 16798 - 1}: Energy
  performance of buildings: Part 1: Indoor environmental input parameters for
  design and assessment of energy performance of buildings addressing indoor
  air quality, thermal environment, lighting and acoustics -- module m1-6
  (07.2015).

\bibitem{Alsen.2016}
N.~Alsen, Energetische und wirtschaftliche bewertung von dezentralen
  ventilatoren in zentralen l{\"u}ftungsanlagen, Phd-thesis, {Universit{\"a}t
  Kassel}, Kassel (2016).

\bibitem{MacQueen.1967}
J.~MacQueen, et~al., Some methods for classification and analysis of
  multivariate observations, in: Proceedings of the fifth Berkeley symposium on
  mathematical statistics and probability, Vol.~1, 1967, pp. 281--297.

\bibitem{VDI2067}
{Verein Deutscher Ingenieure}, {VDI 2067}: Economic efficiency of building
  installations -- fundamentals and economic calculation (2012).

\bibitem{Breuer2024}
J.~H.~P. Breuer, P.~F. Pelz, Efficient and quiet: Optimisation of ventilation
  systems by coupling airflow with acoustics in a multipole approach, in:
  Operations Research Proceedings 2023: Selected Papers of the Annual
  International Conference of the German Operations Research Society (GOR),
  Springer, Accepted in 2023, p. tbd., accepted.

\bibitem{VDI2081}
{Verein Deutscher Ingenieure}, {VDI 2081 Blatt 1}: Air conditioning -- noise
  generation and noise reduction (2022).

\bibitem{WagihNashed.2018}
M.~{Wagih Nashed}, T.~Elnady, M.~{\AA}bom, Modeling of duct acoustics in the
  high frequency range using two-ports, Applied Acoustics 135 (2018) 37--47.
\newblock \href {https://doi.org/10.1016/j.apacoust.2018.01.009}
  {\path{doi:10.1016/j.apacoust.2018.01.009}}.

\bibitem{glav.1997}
R.~Glav, M.~Åbom,
  \href{https://www.sciencedirect.com/science/article/pii/S0022460X96908081}{A
  general formalism for analyzing acoustic 2-port networks}, Journal of Sound
  and Vibration 202~(5) (1997) 739--747.
\newblock \href {https://doi.org/https://doi.org/10.1006/jsvi.1996.0808}
  {\path{doi:https://doi.org/10.1006/jsvi.1996.0808}}.
\newline\urlprefix\url{https://www.sciencedirect.com/science/article/pii/S0022460X96908081}

\bibitem{gijrath2002}
H.~Gijrath, M.~Abom, A matrix formalism for fluid-borne sound in pipe systems,
  in: ASME International Mechanical Engineering Congress and Exposition, Vol.
  36592, 2002, pp. 821--828.

\bibitem{ReynoldsBledsoe1990}
D.~D. Reynolds, J.~M. Bledsoe, Algorithms for HVAC Acoustic, American Society
  of Heating, Refrigerating and Air-Conditioning Engineers (ASHRAE), Atlanta,
  GA, 1990.

\bibitem{suhl2009}
L.~Suhl, T.~Mellouli, Optimierungssysteme: Modelle, Verfahren, Software,
  Anwendungen, Springer-Verlag, 2009.

\bibitem{Carolus.2020}
T.~Carolus, Ventilatoren, {Springer Fachmedien Wiesbaden}, Wiesbaden, 2020.
\newblock \href {https://doi.org/10.1007/978-3-658-29258-4}
  {\path{doi:10.1007/978-3-658-29258-4}}.

\bibitem{eck1972ventilatoren}
B.~Eck, Ventilatoren, Vol.~5, Springer, 1972.

\bibitem{mueller2022}
T.~M. Müller, J.~Neumann, M.~M. Meck, P.~F. Pelz,
  \href{https://www.sciencedirect.com/science/article/pii/S135943112201016X}{Sustainable
  cooling cycles by algorithmically supported design of decentral pump
  systems}, Applied Thermal Engineering 217 (2022) 119084.
\newblock \href
  {https://doi.org/https://doi.org/10.1016/j.applthermaleng.2022.119084}
  {\path{doi:https://doi.org/10.1016/j.applthermaleng.2022.119084}}.
\newline\urlprefix\url{https://www.sciencedirect.com/science/article/pii/S135943112201016X}

\bibitem{madison1949fan}
R.~D. Madison, Fan engineering: an engineer's handbook, on air, its movement
  and distribution in air conditioning, combustion, conveying and other
  applications employing fans, (No Title) (1949).

\bibitem{Rietschel.1994}
H.~Rietschel, Raumklimatechnik, 16th Edition, Springer, Berlin, 1994.

\bibitem{revit}
{Autodesk, Inc.}, {Autodesk Revit},
  \url{https://www.autodesk.com/products/revit/overview}, 2023.1 (2023).

\bibitem{hart2011pyomo}
W.~E. Hart, J.-P. Watson, D.~L. Woodruff, Pyomo: modeling and solving
  mathematical programs in python, Mathematical Programming Computation 3
  (2011) 219--260.

\bibitem{gurobi}
{Gurobi Optimization, LLC}, \href{https://www.gurobi.com}{{Gurobi Optimizer
  Reference Manual}} (2023).
\newline\urlprefix\url{https://www.gurobi.com}

\bibitem{PlanOfWork}
{Royal Institute of British Architects}, {RIBA Plan of Work},
  \url{https://www.architecture.com/knowledge-and-resources/resources-landing-page/riba-plan-of-work},
  access date: 2024-03-22 (2020).

\bibitem{HOAI}
S.~F. Wiesbaden, HOAI 2013-Textausgabe/HOAI 2013-Text Edition: Honorarordnung
  f{\"u}r Architekten und Ingenieure vom 10. Juli 2013/Official Scale of Fees
  for Services by Architects and Engineers dated July 10, 2013,
  Springer-Verlag, 2018.

\bibitem{AIAStandardForm2017}
{American Institute of Architects}, \href{https://www.aiacontracts.com/}{AIA
  Standard Form} (2017).
\newline\urlprefix\url{https://www.aiacontracts.com/}

\bibitem{MissionsMaitriseOeuvre}
\href{https://www.marche-public.fr/Marches-publics/Definitions/Entrees/Elements-mission-maitrise-oeuvre.htm}{Missions
  de maîtrise d’œuvre sur les ouvrages de bâtiment} (1985).
\newline\urlprefix\url{https://www.marche-public.fr/Marches-publics/Definitions/Entrees/Elements-mission-maitrise-oeuvre.htm}

\end{thebibliography}
